\documentclass[12pt]{article}
\usepackage{amsfonts,amsbsy,amssym,psfig}
\newcommand{\del}{\partial}

\newcommand{\x}{{\mathbf{x}}}

\newcommand{\n}{{\mathbf{n}}}

\newcommand{\bi}{{\mathbf{i}}}
\newcommand{\bj}{{\mathbf{j}}}

\newcommand{\M}{\mathbb}

\newcommand{\f}{\frac}

\newcommand{\bb}{\bibitem}
\newcommand{\BF}{\begin{figure}}
\newcommand{\EF}{\end{figure}}
\newcommand{\BE}{\begin{equation}}
\newcommand{\EE}{\end{equation}}
\newcommand{\BEA}{\begin{eqnarray}}
\newcommand{\EEA}{\end{eqnarray}}
\newcommand{\ti}{\textit}

\begin{document}

\title{Are Compact Hyperbolic Models Observationally Ruled Out?}
\author{Kaiki Taro Inoue
\\
Yukawa Institute for Theoretical Physics, 
\\Kyoto University,
Kyoto 606-8502, Japan}
\date{\today}

\maketitle
\begin{abstract}
\noindent
We revisit the observational constraints on compact(closed) hyperbolic(CH) 
models from cosmic microwave background(CMB). 
We carry out Bayesian analyses for CH models
with volume comparable to the cube of the present curvature radius
using the COBE-DMR data and show that 
a slight suppression in the
large-angle temperature correlations owing to the non-trivial topology 
explains rather
naturally the observed anomalously low quadrupole which is
incompatible with the prediction of the standard infinite
Friedmann-Robertson-Walker models. 
While most of positions and 
orientations are ruled out, the likelihoods of CH models 
are found to be much better than those of 
infinite counterparts for some specific positions
and orientations of the observer, leading to less 
stringent constraints on the volume of the manifolds.
Even if the spatial geometry is nearly flat as  
$\Omega_{tot}=0.9-0.95$, suppression of the angular power 
on large angular scales is still prominent for 
CH models with volume much less than the cube of the
present curvature radius if the cosmological constant is dominant at
present.
\end{abstract}

\begin{picture}(0,0)
\put(350,480){
YITP-01-12}
\end{picture}

\section{Introduction}
In the framework of modern cosmology, one often takes it for granted that
the spatial geometry of the universe with finite volume is limited to 
that of a 3-sphere. However, if we drop the assumption of simply 
connectivity of the spatial geometry, then flat and 
hyperbolic(constantly negatively curved) spaces can be closed as
well. Therefore non-positive curvature of the spatial geometry does not 
necessarily implies the infinite space. 
For instance, the simplest example of a constantly curved 
finite-volume space with non-trivial topology 
is a flat 3-torus which can be obtained by identifying the 
opposite faces of a cube by translations in the Euclidean 3-space.
Construction of compact(closed) hyperbolic(CH) spaces is somewhat
complicated but one can systematically create a countably 
infinite number of 
topologically distinct classes of CH spaces by performing topological
\ti{surgeries} on a certain space. 
In order to avoid confusion,  
it might be better to call a cosmological model whose 
spatial geometry is described by a connected and 
simply connected constantly negatively curved space as
an ``infinite hyperbolic'' model rather than an ``open'' model
which has been commonly used in the literature of astrophysics.
\\
\indent
Since 1993, a number of articles concerned with constraints
on the topology of flat models with no cosmological
constant using the COBE-DMR data have been published.
\cite{Sokolov93,Staro93,Stevens93,Oliveira95a,SLS99}
The large-angle temperature 
fluctuations in the cosmic microwave background(CMB)  
discovered by the COBE satellite constrain the 
topological identification scale $L$ \footnote{There is an ambiguity
in the definition of $L$. Here we define $L$ as twice the diameter,
i.e. the longest geodesic distance between arbitrary two points in the
space $M$. Alternatively, one can define $L$ as twice the 
minimum value of the injectivity radius $\textrm{Min}_{p \in M}~
\{ r_{inj}(p) \}$.
Injectivity radius $r_{inj}$ at a point $p$ is equal to the 
radius of the largest connected and simply connected 
ball centered at $p$ which does not cross
itself.} larger than 0.4 times the diameter of the observable region; 
in other words, the
maximum expected number $N$ of copies of the fundamental domain inside 
the last scattering surface in the comoving coordinates 
is $\!\sim$8 for compact flat
models without the cosmological constant\footnote{The constraints are
for models in which the diameter $D$
is comparable to the injectivity radius $r_{inj}$. 
If $D$ is much longer than $r_{inj}$    
then the constraint on $r_{inj}$ is less stringent.\cite{Roukema00}}.
Assuming that the primordial power spectrum is
scale-invariant($n\!=\!1$) then the large-angle 
power is strongly suppressed since fluctuations on scales larger than 
the physical size of the spatial hypersurface 
are strongly suppressed.
\\
\indent
In contrast, a large amount of large-angle fluctuations can be
produced for compact low density models owing to the decay of 
gravitational potential near the present epoch which is known as the 
integrated Sachs-Wolfe effect.\cite{CSS98b} 
If the spatial geometry is
flat or hyperbolic then the physical distance between two separated 
points $P_1$ and $P_2$ which subtends a fixed angle 
at the observation point $O$ 
becomes larger as the distance from $O$ to $P_1$ or $P_2$ increases.
Large-angle fluctuations can be
generated at late epoch when the fluctuation scale ``enters''
the topological identification scale $L$. Recent 
statistical analyses using only the power spectrum have shown that 
the constraints on the topology are not stringent for small
CH models including the smallest (Weeks) and the
second smallest (Thurston) known manifolds and an non-arithmetic orbifold
\cite{CS00,Inoue2,Aurich99,Aurich00} and also for a 
flat compact toroidal model with the cosmological constant.\cite{Inoue5}  
\\
\indent
These results are clearly at odds with the previous constraints
\cite{Bond1,Bond2} on CH models
based on pixel-pixel correlation statistics.
It has been claimed that the statistical analysis using only 
the power spectrum is not sufficient since it can describe only isotropic 
(statistically spherically symmetric) correlations.
This is true inasmuch one considers fluctuations observed at
a particular point. Because any CH spaces are globally
anisotropic, expected fluctuations would be statistically 
globally anisotropic at a particular point. 
\\
\indent
However, in order to constrain CH models, it is also necessary to 
compare the expected fluctuation patterns 
observed at \ti{every place} of the observer to the data
since all CH spaces are globally 
inhomogeneous.\cite{Wolf67} It should be emphasized that the 
constraints obtained in the previous analyses 
are only for CH models
at a \ti{particular observation point} $P$ where the 
injectivity radius is locally maximum 
for 24 particular orientations. $P$ is
rather special one in the sense that some of the mode functions often
(eigenfunctions of the Laplacian) have symmetric structure.\cite{Inoue3}  
It is often the case that the base point $P$ becomes a fixed point of
symmetries of the Dirichlet domain or of the space itself.
At different places the statistically 
averaged temperature correlations significantly changes their 
anisotropic structure as implied by the random Gaussian behavior
of the mode functions of the Laplace-Beltrami operator.\cite{Inoue3}
Thus it is of crucial importance to carry out Bayesian analyses 
incorporating the dependence of the statistically averaged 
correlations on the position of the observer.
\\
\indent
On the other hand, recent CMB observations have succeeded in measuring the
amplitude and the peak of the angular power spectrum on small angular 
scales. Assuming that initial fluctuations are purely adiabatic seeded 
by quantum fluctuations then the first acoustic peak at $l\sim 200$
implies nearly flat geometry.\cite{Boomerang,MAXIMA} 
This seems to deny any 
spherical or hyperbolic models.
However, if one
considers generalized initial fluctuations that include 
isocurvature modes of baryons, cold dark matter and
neutrinos, the uncertainty in estimating the curvature significantly
increases.\cite{BMT00} In order to decrease 
the uncertainty in the cosmological parameters one needs 
information of polarization\cite{BMT00,BMT99} which will be supplied
by the future satellite missions, namely MAP and \ti{Planck}. 
Until then we should interpret the recent CMB observations as which
implies ``not grossly curved spatial geometry'' rather than 
``rigorously flat geometry''.
\\
\indent
Even if the spatial geometry is nearly flat, the effect of the 
non-trivial topology is still prominent provided that 
the spatial hypersurface is much smaller 
than the observable region at present.  
However, we cannot expect a very small CH space since 
there is a lower bound $V_{min}$ for the volume $V$ of CH spaces. 
The precise value of $V_{min}$ is not known but it has been proved
that $V>0.1667R^3$ where $R$ denotes the curvature radius for 
CH manifolds. The smallest known example has been discovered 
by Weeks\cite{Weeks85} 
which has volume $0.9427R^3$ called the Weeks manifold. 
If we allow CH orbifold models then the volume can be much smaller.
For instance, the smallest known orbifold has volume
$V=0.0391R^3$.
\\
\indent
Suppose that the cosmological constant or
the ``missing energy'' with negative equation-of-state dominates the 
present energy density as the recent observations of SNIa imply
\cite{Perlmutter,Riess} then the comoving radius of 
the last scattering surface in 
unit of curvature radius becomes large because of the slow decrease in 
the cosmic expansion rate in the past.
For instance, the number $N$ of 
copies of the fundamental domain inside
the last scattering in the comoving coordinates is approximately 17.2
for a Weeks model with $\Omega_\Lambda\!=\!0.65$ and $\Omega_m\!=0.2\!$ 
whereas $N\!=\!2.5$ if $\Omega_\Lambda\!=\!0$ and
$\Omega_m\!=0.85\!$. $N$ can be much larger for small orbifold
models. Thus, even in the case of nearly flat geometry, there are 
possibilities that we can observe the imprint of the non-trivial topology.
\\
\indent
In this paper, we investigate the CMB anisotropy in CH models(manifold
and orbifold) with small volume with or without the 
cosmological constant. In section 2
we briefly describe the necessary ingredients of the geometry 
and topology of CH spaces. In section 3 the time evolution
of the scalar-type perturbation which can be used for simulating 
the CMB anisotropy is described. 
In section 4 we estimate the degree 
of suppression in the large-angle power for manifold and orbifold
models. In section 5 we carry 
out Bayesian analyses using the COBE-DMR data. In the last section 
we summarize our results.

\section{Compact Hyperbolic Spaces}
The discrete subgroup $\Gamma$ of 
the orientation-preserving 
isometry group of the simply-connected hyperbolic 3-space $H^3$
(which is isomorphic to $PSL(2,\M{C})$) is called the Kleinian group. 
Any orientable CH 3-spaces (either manifold or
orbifold) can be described as compact quotients\footnote{The 
orientation-reversing isometry 
group of $H^3$ is isomorphic to
$PSL(2,{\M{C}})\bj$. If $\Gamma$ includes an element which is isomorphic to 
the discrete subgroup of $PSL(2,{\M{C}})\bj$ then the compact quotient 
$M$ is unorientable.}$M=H^3/
\Gamma$. 
Let us represent $H^3$ as an upper half space
($x_1,x_2,x_3$) for which the metric is given by
\BE
ds^2=\f{R^2(dx_1^2+dx_2^2+dx_3^2)}{x_3^2},
\EE
where $R$ is the curvature radius. In what follows,
$R$ is set to unity without loss of generality.
If we represent a point $p$ on the upper-half space, as a quaternion whose
fourth component equals zero, then the actions of  $PSL(2,{\M
C})$ on $H^3 \cup {\M C} \cup\{ \infty\} $ take the form  
\begin{equation}
\gamma: p\rightarrow p'=\f{a p+ b}{c p+ d}, 
~~~~~ad-bc=1,~~~p\equiv z + x_3 {\bj}, ~~~z=x_1+x_2 {\bi},
\end{equation}
where a, b, c and d are complex numbers and $1$, $\bi$ and $\bj$ are
the components of the basis of quaternions.
The action $\gamma$ is
explicitly written as   
\begin{eqnarray}
\gamma:H^3\cup{\M C}\cup\{ \infty \} &\rightarrow& 
H^3\cup{\M C}\cup\{ \infty \},
\nonumber
\\
\nonumber
\\
\gamma:(z(x_1,x_2),x_3)~~ &\rightarrow& 
\Biggl
( \f{(az+b)(\overline{cz+d})+a\bar{c}x_3^2}{|cz+d|^2+|c|^2x_3^2},
\f{x_3}{|cz+d|^2+|c|^2x_3^2}\Biggr), 
\end{eqnarray}
where a bar denotes a complex conjugate.
Elements of $\Gamma$ for orientable CH manifolds
are $SL(2,\M{C})$ conjugate to 
\BE
\pm \left(
\begin{array}{@{\,}cc@{\,}}
\exp({l/2+i \phi/2}) & 0\\ 
0 & \exp({-l/2-i \phi/2})
\end{array}
\right ),~0<\phi\le 2\pi,~l>0
\EE
which are called \ti{loxodromic} if $\phi\ne0$ and
\ti{hyperbolic} if $\phi=0$. 
\\
\indent
There are other classes of isometries of $H^3$ which have fixed
points, namely, \ti{parabolic} and \ti{elliptic} elements.
In matrix representation, they are $SL(2,\M{C})$ conjugate to 
\BE
\pm \left(
\begin{array}{@{\,}cc@{\,}}
1 & 1\\ 
0 & 1
\end{array}
\right ),~~\ti{parabolic}
\EE
and
\BE
\pm \left(
\begin{array}{@{\,}cc@{\,}}
\exp({i \phi/2}) & 0\\ 
0 & \exp({i \phi/2})
\end{array}
\right ),~0<\phi\le 2\pi,~~\ti{elliptic}
\EE
, respectively. An orientable CH orbifold can 
be realized as a compact quotient
$H^3$ by $\Gamma$ which includes an elliptic 
element. If $\gamma$ includes a parabolic element, then the quotient space
is called a cusped hyperbolic manifold. A cusp point  
corresponds to a fixed point at infinity 
by the parabolic transformation.
In what follows we denote a ``cusped manifold'' as a finite-volume
hyperbolic manifold with one cusp point or more.
\\
\indent
Topological construction of CH spaces starts with a cusped manifold
$M_c$. Let us first consider the two-dimensional case. 
If we glue two ideal hyperbolic triangles 
together on each sides, then we have a thrice punctured sphere. 
Because each side is isometric to a line, we can modify the gluing map
by an arbitrary translation (in this case it is parametrized by ${{}\M{R}}^3$).
However, the obtained surface is not always complete. Let $h$ be a
horocycle\footnote{Horocycles centered at a vertex
$v$ at infinity are the curves orthogonal to all lines through
$v$.}segment, that is orthogonal to the two sides of the triangle
centered at a vertex $v$. Extend $h$ counterclockwise about $v$ by
horocycle segments that meet successive edges of ideal triangles.  
Finally it re-enters the orthogonal triangle as a horocycle concentric
with $h$, at a distance $d(v)$. The obtained surface is complete 
if and only if $d(v)=0$ for all vertices $v$\cite{ThurstonLN}.
In this case, the condition $d(v)=0$ yields a hyperbolic manifold with 
three cusp points where the horocycles correspond to periodic geodesics
associated with parabolic elements.  
\\
\indent
\BF[t]
\centerline{\psfig{figure=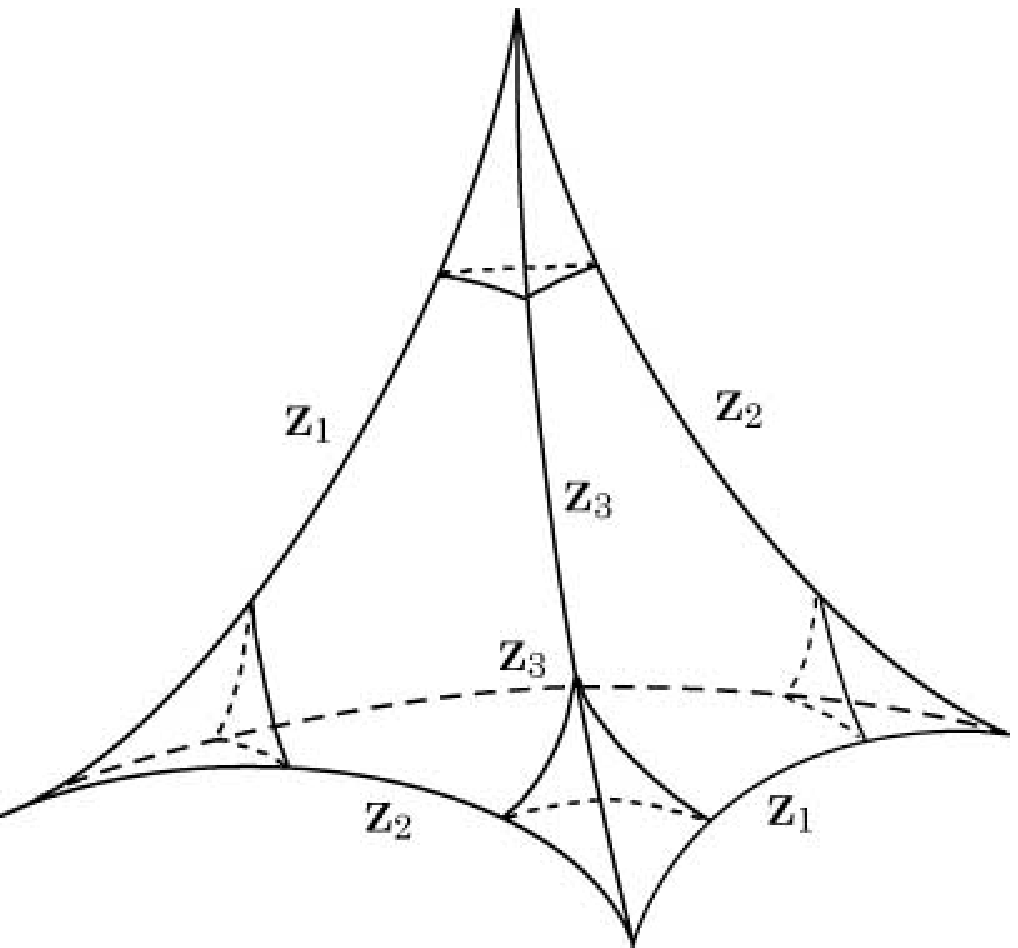,width=9cm}}
\caption{Parameterization of an ideal tetrahedron. Each edge $e$ is
labelled with a complex number $z(e)$ and opposite edges have the same 
label. The four links (=Euclidean triangles realized as the
intersections with horospheres in the neighborhood of vertices) 
are congruent each other. An ideal tetrahedron up to orientation-preserving 
similarity can be specified by one
link which is parametrized by one of the $z(e)$.}
\label{fig:idealtetrahedron}
\EF
Similarly, one can construct a cusped 3-manifold
by gluing ideal polyhedra. Here we consider gluing of a set of
$n$ ideal tetrahedra $T_1, ... ,T_n$. $T_i$ is determined by the 
three dihedral angles $\alpha,\beta$ and $\gamma$ which satisfy
$\alpha+\beta-\gamma=\pi$. Because the dihedral angles 
of opposite edges are equal, $T_i$ is essentially 
determined by two parameters. In order to represent the shape
of an ideal tetrahedron $T_i$ , one can use the \ti{link} $L(v)$ of a
vertex $v$ at infinity which is an Euclidean triangle realized as the 
intersection of a horosphere\footnote{Horospheres centered at a vertex
$v$ at infinity are the surfaces orthogonal to all lines through $v$.}
with $T_i$ in the neighborhood of $v$. $L(v)$ can be 
described by two parameters and 
determines $T_i$ up to congruence. 
In order to parameterize $L(v)$, it is convenient to represent the 
Euclidean plane $E^2$ as
$\M{C}$.\cite{ThurstonLN} To each vertex $v_1$ of a triangle ($v_1,v_2,
v_3$) (the vertices are labeled in a clockwise order)in $\M{C}$ 
we associate the ratio 
\BE
z(v_1)=\f{v_3-v_1}{v_2-v_1}
\EE
of the sides adjacent to $v_1$. The ratios(with any starting point) 
should satisfy
\BEA
z(v_1)z(v_2)z(v_3)&=&-1,
\nonumber
\\
1-z(v_1)+z(v_1)z(v_2)&=&0.
\EEA
Thus any $z(v_i)$ determines the other two $z(v_j)$s. 
An ideal tetrahedron can be specified by any $z(v_i)$ having
two parameters up to orientation-preserving similarity (figure 
\ref{fig:idealtetrahedron}). 
The gluing conditions of ideal tetrahedra $T_1, ... ,T_n$
are algebraically given by
\BEA
z(e_1)\cdot z(e_2)\cdot \cdots \cdot z(e_m)&=&1,
\nonumber
\\
\textrm{arg} z(e_1)+\cdots+\textrm{arg} z(e_m)&=&2\pi,~0\le 
\textrm{arg}\le\pi
\EEA
where $e_i$s are the edges of ideal tetrahedra.
Note that even if the above conditions are satisfied the obtained
manifold is sometimes incomplete where the developing image of
$L(v)$ is ${\M{C}}-0$ after an appropriate translation. 
If $L(v)$ properly tessellates ${\M{C}}$ then 
the obtained manifold $M$ is complete and the boundary of the 
neighborhood of a vertex at infinity is either a torus or a Klein bottle. 
$M=M_c$ is topologically equivalent 
to the complement of a knot $K$ or link $L$(which consists of knots)
in a 3-sphere $S^3$ or some other closed 3-space. 
A surgery in which 
one removes the tubular neighborhood $N$ of $K$
whose boundary is homeomorphic to a torus, and replace $N$
by a solid torus so that a meridian\footnote{Given 
a set of generators $a$ and $b$
for the fundamental group of a torus, a closed curve $C$  
which connects a point $x$ in the torus with $ax$ is 
called a {\ti{meridian}} and
another curve which connects a point $x$ with $bx$ is called a
{\ti{longitude}}.}
in the solid torus
goes to $(p,q)$ curve\footnote{If a closed curve $C$ connects 
a point $x$ with another point $(pa+qb)x$ where $p$ and $q$ 
are integer, $C$ is called a $(p,q)$ curve.}
on $N$ is called $(p,q)$ \ti{Dehn surgery}. Except for a finite number
of cases, Dehn surgeries on $K$ where $p$ and $q$ are co-prime 
always yield CH 3-manifolds which
implies that most compact 3-manifolds are hyperbolic\cite{Thurston82}
since every closed 3-manifolds can be obtained by performing such
Dehn surgeries. CH 3-orbifolds whose singular sets consist 
of lines can be obtained by $(p,q)$ Dehn surgeries 
where $p$ and $q$ are not co-prime.
\\
\indent
The above procedure for obtaining CH 3-spaces is manifestly defined
but somewhat complicated for computation by hand. A computer program
``SnapPea'' developed by Weeks\cite{SnapPea} can numerically 
handle the procedure, namely computation of hyperbolic 
structure of cusped manifolds, performing 
various Dehn surgeries and so on.  SnapPea can also compute the volume,
fundamental group, homology, symmetry group, Chern-Simon invariant, 
length spectrum and other quantities. 
\BF
\centerline{\psfig{figure=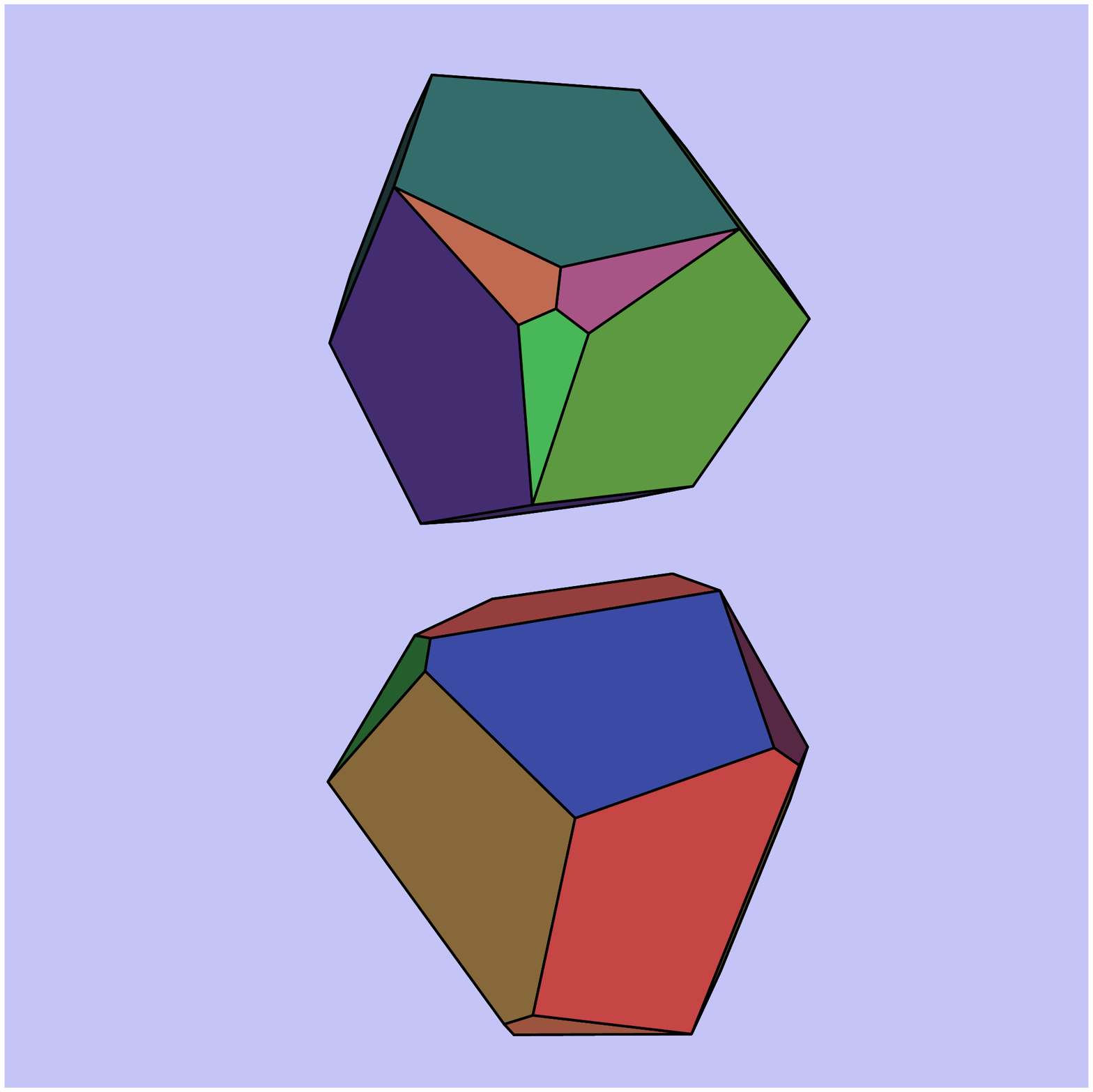,width=6.5cm}
\psfig{figure=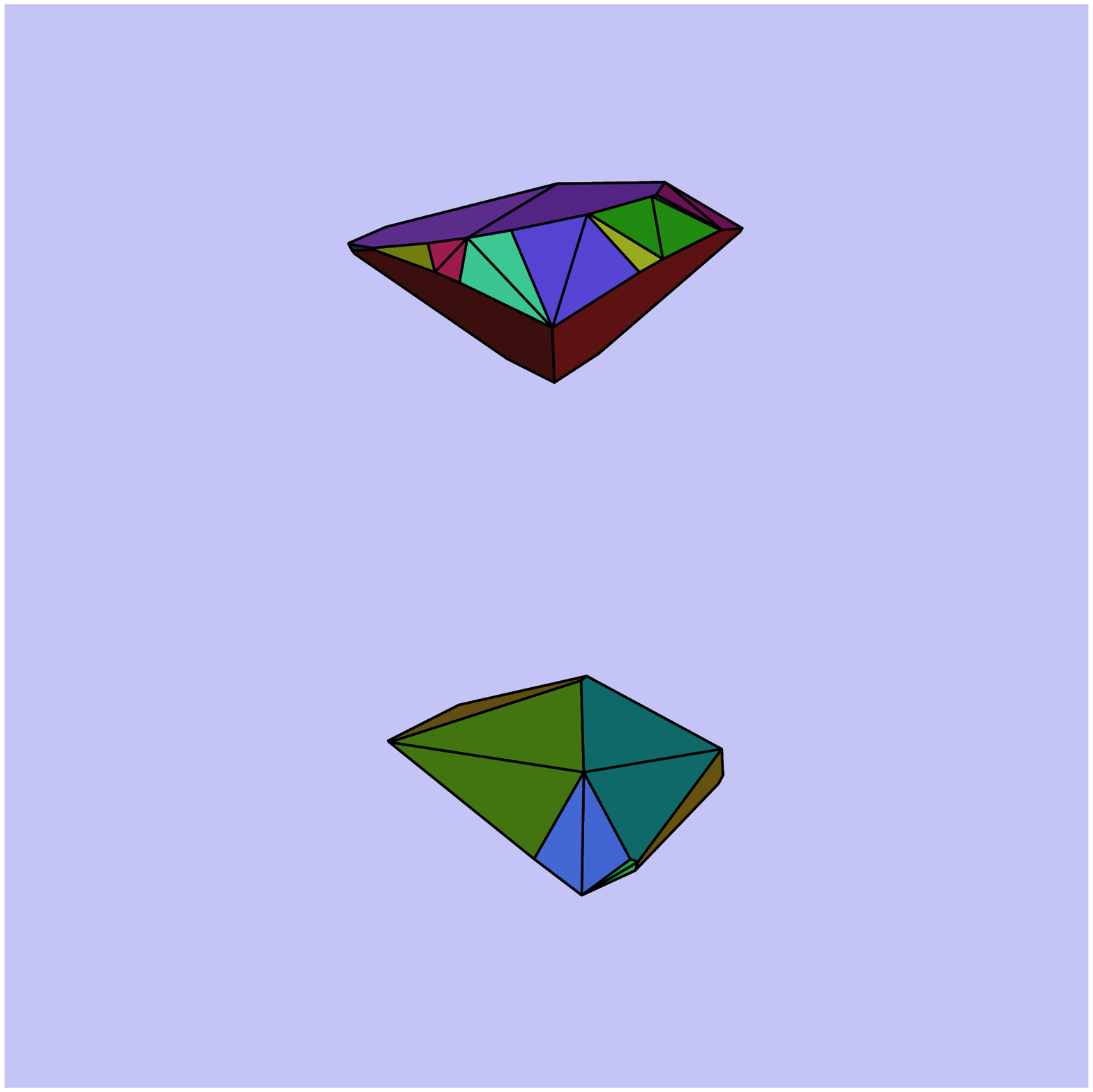,width=6.5cm}
}
\caption{The Dirichlet fundamental domains 
of the Weeks manifold (volume$=0.9427$)(left) and the smallest
orbifold that is covered by
the Weeks manifold (volume$=0.0786$)(right)    
viewed from opposite directions in the
Klein(projective) coordinates. Colors on the faces represent the
identification maps. }
\label{fig:Weeks}
\EF
\\
\indent
In contrast to compact flat spaces with no scale,
there is a lower bound for the volume of CH 3-spaces. 
It has been proved that the volumes of CH 3-manifolds 
cannot be smaller 
than $0.16668\dots$\cite{GMT96} although no concrete example
with such volume is known.   
The smallest known CH 3-manifold is called the Weeks manifold with
volume $0.9427$ which can be obtained by a (3,-1) Dehn surgery 
on a cusped manifold $m003$ with volume 2.0299(one of the smallest known
cusped manifold). Performing a (-2,3) Dehn surgery on
m003 yields the second smallest known manifold called the Thurston manifold
with volume $0.9814$. Except for a finite number of cases, Dehn
surgeries on m003 yield CH manifolds with volume less than that of 
m003.\cite{Thurston82} 
As $|p|$ and $|q|$ becomes large, the volume of the space
converges to that of the original cusped manifold $M_c$.   
Therefore, for the purpose of classification of CH spaces it is better to
exclude ``almost non-compact'' spaces  
that are very similar to $M_c$.\footnote{The Hodgson-Weeks census
consists of the data of 11,031 CH manifolds with length of the shortest
periodic geodesic $l>0.3$ and volume 
$V<6.4535$.}
The volume of CH 3-orbifolds can be much smaller since 
every CH 3-orbifold is covered by a CH 3-manifold(there are an 
infinite number of coverings).
For instance, the smallest
volume of a CH 3-orbifold whose finite-sheeted cover is the Weeks
manifold is $V=0.9427/12=0.0786$ (see figure \ref{fig:Weeks})since 
the isometry group of the Weeks manifold is $D6$(dihedral group),
which does not have any fixed planes. 
The smallest known orientable
3-orbifold has volume $V=0.0391$ which also does not have any 
fixed planes.\cite{Chinburg86}
A known lower bound for the volume of orientable orbifolds 
is $V>0.0000013$.\cite{Meyerhoff88} 
\section{CMB Anisotropy}
In what follows we assume that the matter consists of two
components:a relativistic and an non-relativistic one
($\Omega_m=\Omega_r+\Omega_n$). From the Friedmann equation, the
integral representation for the time evolution of the 
scale factor $a$ (normalized to 1 at present time) in terms of the 
conformal time $\eta$ is given by 
\BE
\eta(a)=R \int_0^a 
\f{da}{\sqrt{\textrm{sign} K((\Omega_{n}a+\Omega_{r}+\Omega_\Lambda a^4)/
(\Omega_{m}+\Omega_\Lambda-1)-a^2)}}, K\ne0,
\label{eq:bg}
\EE
where $R=|K|^{-1/2}$ is the curvature radius at
present. 
The above presentations can also be written in terms of some special
functions but they are too complex to describe here. 
\\
\indent
The scalar perturbation equations are exactly same as those without 
the cosmological constant. Let us consider the adiabatic case and 
assume that the anisotropic pressure is negligible.
Then from the linear perturbation equation, the time evolution of the 
Newtonian curvature perturbation $\Phi$ of $k$ mode is given by,
\cite{Mukhanov92}
\BE
\Phi''+3 {\cal{H}}(1+c_s^2)\Phi'+c_s^2k^2\Phi+
(2 {\cal{H}}'+(1+3 c_s^2)({\cal{H}}^2-K))\Phi=0,
\label{eq:phi-evolution}
\EE
where $'$ denotes the conformal time derivative $d/d\eta$, 
${\cal{H}}\equiv a'/a$, G is the Newton's constant, 
$c_s$ is the sound speed of the
fluid which is explicitly written in terms of the density
of relativistic matter $\rho_r$ and non-relativistic matter $\rho_m$  
\BE
c_s^2=\f{1}{3} \f{\rho_r}{\rho_r+\f{3}{4}\rho_m}. 
\EE
Let us consider the early universe when the radiation is dominant
and the effect of the curvature and the cosmological constant is
negligible. Then (\ref{eq:bg}) yield the evolution
of the scale factor as $a \propto \eta$ and
(\ref{eq:phi-evolution}) is reduced to
\BE
\Phi''+\f{4}{\eta}\Phi'+\f{k^2}{3}\Phi=0,
\EE 
which has an analytic solution
\BE
\Phi(\eta)=\eta^{-3}\{(\omega \eta \cos(\omega \eta)
-\sin(\omega \eta))C_1+(\omega \eta \sin(\omega \eta)
-\cos(\omega \eta))C_2\},
\label{eq:analytical}
\EE
where $\omega=k/\sqrt{3}$ and $C_1$ and $C_2$ are constants which
depend on $k$.\cite{Mukhanov92} Around $\eta=0$, 
(\ref{eq:analytical}) can be expanded as
\BE
\Phi(\eta)=\eta^{-3}\{C_2+\f{C_2}{2}\omega^2\eta^2-\f{C_1}{3}\omega^3\eta^3
-\f{C_2}{8}\omega^4 \eta^4+{\cal O}((\omega \eta)^5) \}.
\label{eq:expansion}
\EE
Thus for the long-wavelength modes($\omega \eta <<1$),
the non-decaying mode ($C_2=0$) has constant amplitude 
and the time derivative at the initial time vanishes,
\BE
\Phi\sim \textrm{const},~~~~~\Phi'(0)=0.
\EE
During the radiation-matter equality time the 
equation of state change from $w=1/3$ to 
$w=0$. Then the amplitude of long-wavelength modes 
changes by a factor of $9/10$. In the curvature or $\Lambda$ 
dominant epoch the amplitude decays as $1/a$. 
\\
\indent
Assuming that the matter is dominant at the last scattering
\footnote{For low density models, the ordinary Sachs-Wolfe term in 
(\ref{eq:SW}) may not be valid since the last scattering may occur before
full-matter domination. 
However, on large angular scales (\ref{eq:SW}) still gives a 
good approximation since 
the integrated Sachs-Wolfe term 
dominates over the ordinary Sachs-Wolfe term in low density models.} 
and the anisotropic pressure is negligible,
the temperature fluctuation on large angular scales can be 
written in terms of the Newtonian curvature perturbation
$\Phi$ as 
\BE
\f{\Delta T}{T}(\n)=-\f{1}{3} \Phi(\eta_\ast, (\eta_0-\eta_\ast)\n)
-2\int _{\eta_\ast}^{\eta_0} \f {\del \Phi(\eta, (\eta_0-\eta)\n)}
{\del \eta} d \eta,
\label{eq:SW}
\EE
where $\n$ denotes the unit vector which points the sky
and $\eta_\ast$ and $\eta_0$ correspond to the last scattering
conformal time and the present conformal time, 
respectively.\cite{SachsWolfe,HSS} 
The first term in the right-hand 
side in (\ref{eq:SW}) describes the ordinary Sachs-Wolfe (OSW)
effect while the second term
describes the integrated Sachs-Wolfe (ISW) effect. 
\section{Angular power spectra}
In order to analyze the temperature
anisotropy in the sky it is 
convenient to expand it in terms of spherical harmonics
$Y_{lm}$ as
\BE
\f{\Delta T}{T}(\n)=\sum_{lm} a_{lm} Y_{lm}(\n),
\EE
where $\n$ is the unit vector along the line of sight.
Because each mode of scalar perturbation 
evolves independently in the locally isotropic and homogeneous
background space, we need only the information of eigenmodes $u_\nu$
of the Laplace-Beltrami operator $\Delta$ in the CH space (i.e. 
elements of an orthonormal basis of $L^2(H^3/\Gamma)$)   
which can be expanded in terms of 
eigenmodes in the simply connected infinite hyperbolic space
in the spherical coordinates ($\chi,\theta,\phi$)
as
\BEA
u_\nu&=&\sum_{l m} \xi_{\nu l m}\,X_{\nu l}(\chi) Y_{l m}(\theta,\phi),
\nonumber
\\
X_{\nu l}(\chi)
&=&\f{\Gamma  (l+1+\nu i)}{\Gamma (\nu i)} \sqrt{\f{1}{\sinh \chi}}
P^{-l-1/2}_{\nu i-1/2}(\cosh
\chi),
\label{eq:u}
\EEA 
where $~\nu^2=k^2-1$, $Y_{l m}$ denotes (complex) 
spherical harmonics, $P$ is the associated Legendre function 
and $\xi_{\nu l m}$ represents the expansion coefficients.
\\
\indent
Assuming that the initial fluctuations obey the Gaussian statistic, 
and neglecting the tensor-type perturbations and anisotropic pressure 
and matter is dominant at the last scattering epoch,
the angular power spectrum $C_l=<|a_{lm}|^2>$ ($<>$ denotes an 
ensemble average over the initial perturbation) for CH models 
can be written as
\BEA
(2\,l+1)\,C_l
&=&\sum_{\nu,m}\f{4 \pi^4~{\cal P}_\Phi(\nu) }
{\nu(\nu^2+1)V(M)}~|\xi_{\nu l m}|^2 |F_{\nu l}|^2 ,
\nonumber
\\
F_{\nu l}(\eta_o)
\!\! &\equiv& \!\!\f{1}{3}
\Phi_t(\eta_\ast) X_{\nu l}(\eta_o\!-\!\eta_\ast)
\!+\!\! 2 \!\!\int_{\eta_\ast}^
{\eta_o}\!\!\!\!\!\!d \eta\, 
\f{d\Phi_t}{d \eta}X_{\nu l}(\eta_o\!-\!\eta),\label{eq:CMBcorCH}
\EEA
where $\nu\!=\!\sqrt{k^2-1}$, ${\cal P}_\Phi(\nu) $ 
is the initial power spectrum, and
$\eta_\ast$ and $\eta_o$ are the
last scattering and the present conformal time, respectively. 
$\Phi_t=\Phi/\Phi(0)$ and $V(M)$ is the comoving volume of the 
space. 
The (extended) Harrison-Zel'dovich spectrum corresponds to 
${\cal P}_\Phi(\nu)\!=\!Const.$
which we shall use as the initial 
condition. It should be noted that $C_l$'s depend on the 
position and orientation of the observer since CH spaces are
globally inhomogeneous. Therefore, it is
better to consider the ensemble average
$\hat{C}_l\equiv<C_l>$ taken over the position of the observer.
Using Weyl's asymptotic formula 
one can easily show that $\hat{C}_l$'s coincide with those for the 
infinite counterpart in the limit $\nu \rightarrow \infty$
provided that $<|\xi_{\nu l m}|^2>\propto
\nu^{-2}$.
\footnote{
Although we have assumed that the 
normalization factor in $<|\xi_{\nu l m}|^2>$
does not depend on the volume of the space, the volume factor in 
(\ref{eq:CMBcorCH}) might
not be necessary if $<|\xi_{\nu l m}|^2>$
is proportional to $1/V(M)$.}
Thus all we have to do is to numerically compute the expansion coefficients 
$\xi_{\nu l m}$ which contain the information of global topology and 
geometry.\footnote{The low-lying eigenmodes of the two smallest 
manifolds (Weeks and Thurston) have been computed using 
the direct boundary element method.\cite{Inoue1} For other small CH manifolds
the eigenvalues have been computed using the periodic 
orbit sum method.\cite{Inoue4}}
\\
\indent
In order to estimate the 
suppression in the angular power $\hat{C}_l$ it is convenient 
to write $\hat{C}_l$ in terms of the transfer function $T_l(k)$
which satisfies \cite{HueD}
\BE
\f{2l+1}{4 \pi}\tilde{C}_l= \int T_l^2(k){\cal P}(k) \f{dk}{k},
\EE
where $\tilde{C}_l$ denotes the angular power for the infinite
counterpart. For CH models, $\hat{C}_l$ is written as a sum
of $T_l^2(k)$ for discrete wave numbers $k_i$.
\\
\indent
On large angular scales, the behavior of the angular power 
$\hat{C}_l$ is determined
by low-lying modes which are susceptible to the global 
topology of the background geometry 
since the amplitude of each mode is proportional to 
$\sim \nu^{-5} |F_{\nu l}|^2$.   
As shown in figure \ref{fig:TL5CHM}, one can see that 
the transfer function corresponding to
the first eigenmode $T_l^2(k_1)$ mimics 
the angular dependence of $\hat{C}_l$.
\\
\indent
The large-angle power owing to the OSW effect 
suffers a significant suppression since fluctuations
on scales beyond the actual size of the space are not allowed.
The angular cutoff scale $l_{cut}$ above which the OSW contribution
suffers a suppression corresponds to the angular scale of 
the first eigenmode on the last scattering surface, 
which can be written in terms of the comoving volume $V(M)$
and the smallest non-zero wavenumber $k_1$ as \cite{Inoue4}
\BE
l_{cut}=\f{k_1}{4} \sqrt{V^{-1}((\sinh(2(R_{LSS}+r_{ave}))
-\sinh(2(R_{LSS}-r_{ave}))-4 r_{ave}))},
\EE
where $V=\pi(\sinh(2r_{ave})-2r_{ave})$ and $R_{LSS}$ denotes the comoving
radius of the last scattering surface. For the Weeks models
($V=0.94$ and $k_1=5.27$), $l_{cut}=26$ for $\Omega_0=0.2$ and
$l_{cut}=7$ for $\Omega_0=0.6$. 
\BF[tp]
\centerline{\psfig{figure=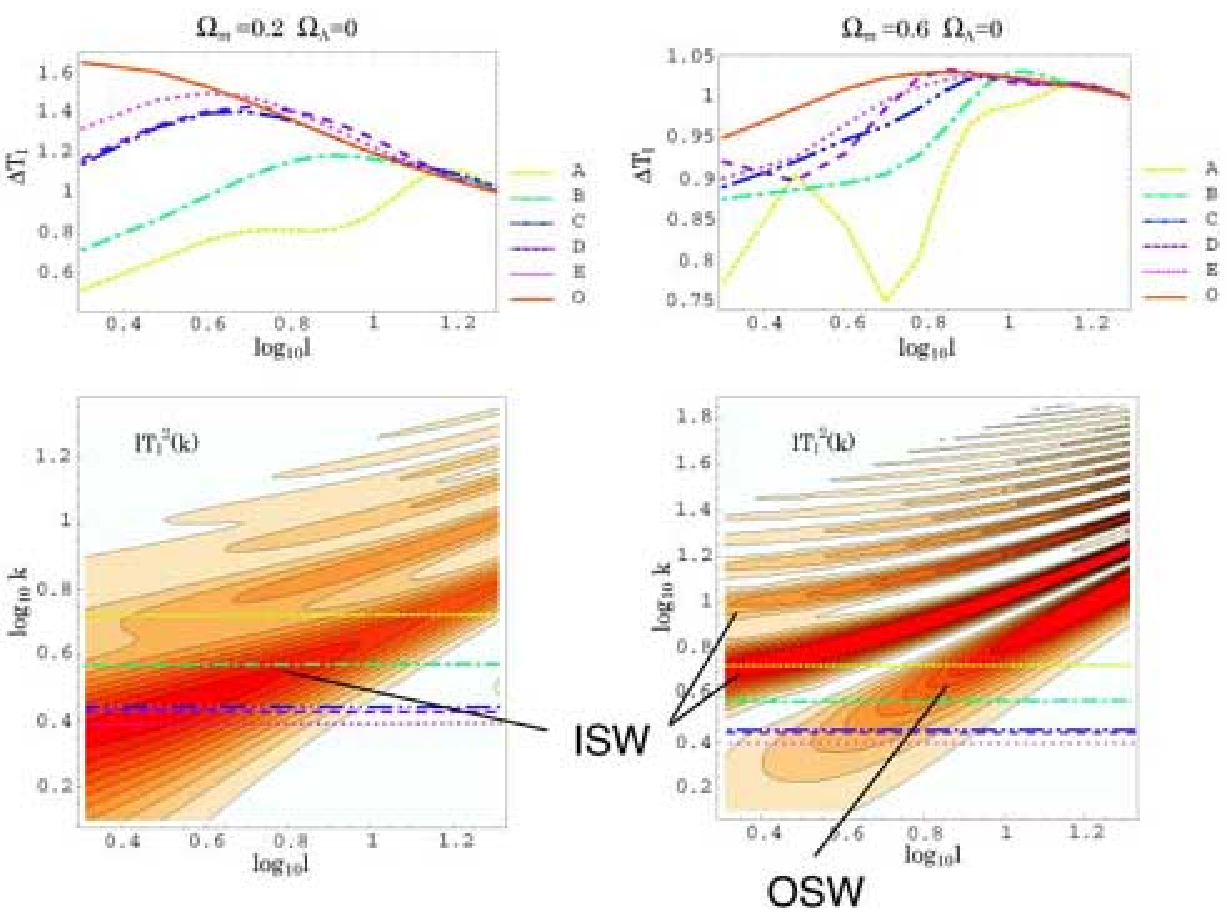,width=18.5cm}}
\caption{Suppression in the large-angle power 
$\Delta T_l\equiv \sqrt{l(l+1)\hat{C}_l/(2
\pi)}$ for 5 CH models (name,volume)=
A:(m003(3,-1)(Weeks),0.94), B:(m010(-1,3),1.9),
C:(m082(-2,3),2.9), D:(m288(-5,1),3.9) E:(s873(-4,1), 4.9)
in comparison with the one for the infinite hyperbolic model(denoted
as O). 
All the plotted values are normalized by 
$\Delta T_{20}$ for the infinite hyperbolic model.
The $l$ dependence of $\Delta T_l$ can be approximately given by
the transfer function $lT^2_l(k_1)$. 
First eigenvalues $k_1$ for the five models are
represented as horizontal lines in lower figures. 
The unit of $k$ is equal to the inverse of the present
curvature radius.}
\label{fig:TL5CHM}
\EF
From figure \ref{fig:TL5CHM} one can
see that the angular scales 
$l_{cut}$ corresponding to the 
intersection of $k=k_1$ and the OSW 
ridge of the transfer function well agree with these 
analytic estimates. On smaller angular scales $l>>l_{cut}$,
the powers asymptotically converge to those for the infinite
counterpart.
\\
\indent
Because the large-angle power in the COBE-DMR data is 
nearly flat, it seems that the suppression leads to a bad fit to
the data on large angular scales.
However, for low density models we should consider 
the ISW effect owing to the gravitational potential decay 
at the curvature dominant epoch $1+z\sim (1-\Omega_0)/\Omega_0$
or $\Lambda$ dominant epoch $1+z\sim (\Omega_\Lambda/\Omega_0)^{1/3}$
well after the last scattering time.
For a given angular scale $l$ 
the comoving scale $k^{-1}$ of a fluctuation that is produced at 
late time is decreased. On the other hand, 
fluctuations do not suffer suppression if the comoving scales of
fluctuations are sufficiently smaller than the actual size of the space
$k^{-1}<<r_{ave}$. Therefore, the suppression on the ISW contribution 
is less stringent compared with the OSW contribution.
Interestingly, the suppression on the angular power 
owing to the mode-cutoff reduces the excess power 
owing to the ISW effect, resulting in a nearly flat power with 
a slight suppression on large angular scales $2\le l \le 10$. 
\\
\indent
The suppression of the power is crudely estimated by the number $N$ of 
the copies of the fundamental domain inside the last scattering
surface in comoving coordinates. 
Suppose that $\Omega_\Lambda\!=\!0$. Then the comoving radius of the
last scattering surface in terms of the present curvature radius 
$R$ 
\BE
R_{LSS}\approx R\cosh^{-1}(2/\Omega_m-1),
\EE
gives the comoving volume $v_{LSS}$ 
of the ball inside the last scattering surface 
\BE
v_{LSS} \approx \pi R^3(\sinh (2 R_{LSS}/R)-2 R_{LSS}/R).
\EE
For example, $v_{LSS}\!\sim\!490 R^3$ for 
$(\Omega_m,\Omega_\Lambda)\!=\!(0.2,0)$ 
whereas $v_{LSS}\!\sim\! R^3$ for 
$(\Omega_m,\Omega_\Lambda)\!=\!(0.9,0)$. Thus in nearly flat
models, the imprint of the non-trivial topology is prominent only 
for the case where the volume is smaller than $R^3$.
However, if one 
includes the cosmological constant then $R_{LSS}$ (in unit of $R$) 
becomes large because of a slow increase in 
the cosmic expansion rate in the past. For instance,
$N=8.7$ for a Weeks model (the smallest known manifold) with 
$\Omega_\Lambda\!=\!0.7$ and $\Omega_m\!=0.2\!$ 
whereas $N\!=\!1.2$ for 
$\Omega_\Lambda\!=\!0$ and $\Omega_m\!=0.9\!$. 
Furthermore, if one allows orbifold models,
$N$ can be much larger than these values. 
For instance, 
$N=51.8$ for the smallest orbifold (volume$=0.0391R^3$)
with $\Omega_\Lambda\!=\!0.7$ and $\Omega_m\!=0.25\!$. 
\BF[t]
\centerline{\psfig{figure=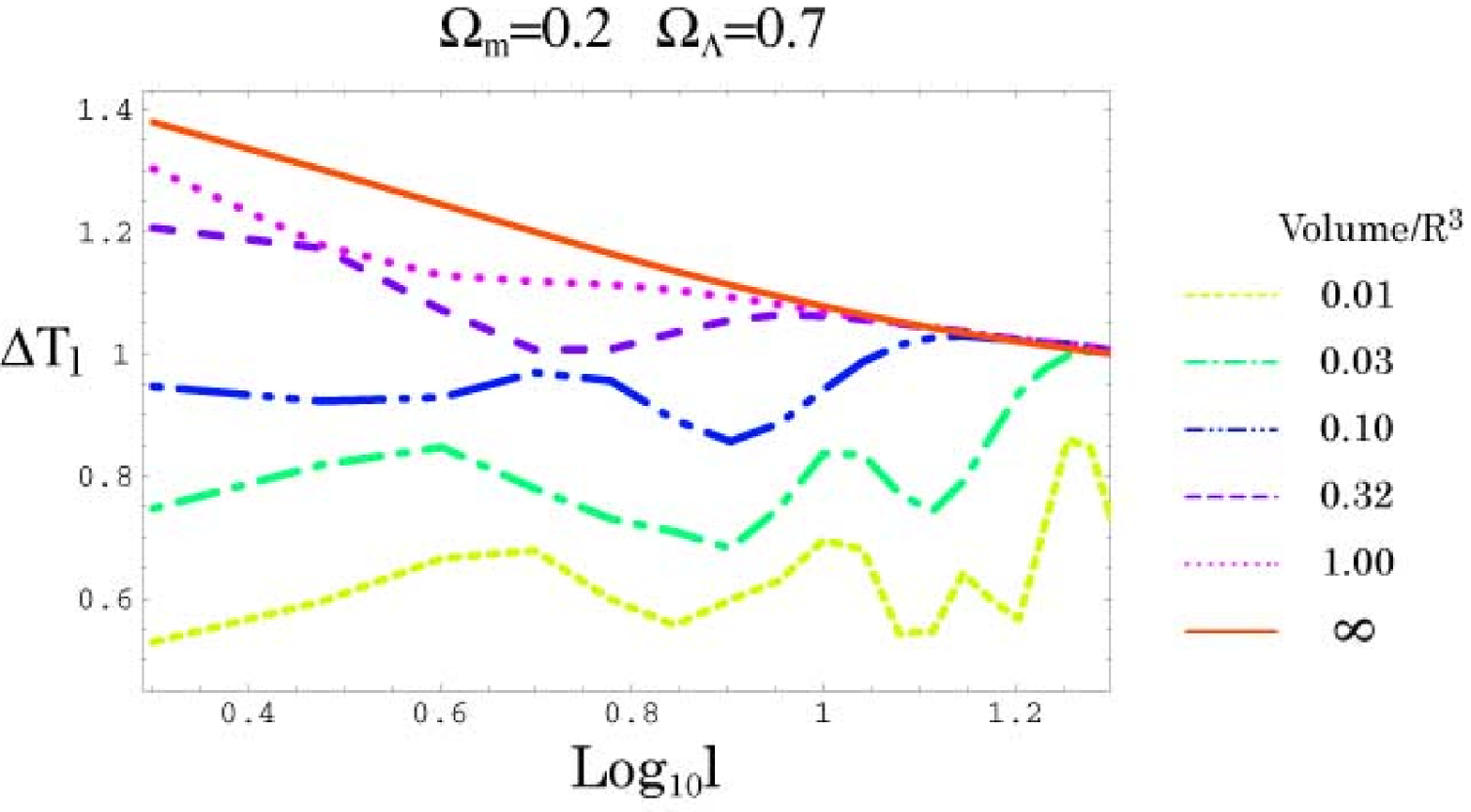,width=14cm}}
\caption{Suppression in large-angle power $\Delta
T_l=\sqrt{l(l+1)\hat{C}_l/2 \pi}$ for small (hypothetical) orbifold models
with $\Omega_m=0.2$ and $\Omega_\Lambda=0.7$ in comparison with the
infinite hyperbolic model with the same density parameters. The
amplitude is normalized by $\Delta T_{20}$ for the infinite hyperbolic model.
}
\label{fig:TLorbifolds}
\EF
\\
\indent
Now let us consider the angular power spectrum for orbifold models. 
Assuming no very short periodic geodesics and no supercurvature 
modes, one can crudely estimate the 
low-lying eigenvalues \cite{Inoue4} from the 
asymptotic form of the number function(spectral staircase) \cite{AM96}
\BE
n(E)=\f{V(M)}{6 \pi ^2}\nu^3+c_2 \nu^2+c_1 \nu + c_0 + O(e^{-\pi
\nu/5}), ~\nu^2=E-1,
\label{eq:Weyl}
\EE
where $n(E)$ denotes the number of eigenvalues of the Laplace-Beltrami 
operator equal to or less than $E$ and $c_i$'s are the constants
determined from the plane reflection, elliptic and inverse elliptic
elements of the discrete isometry group 
and the area of the fixed planes.  For orientable orbifolds
with only fixed lines, the constants vanish except for $c_1$
which is written in terms of elliptic elements. If we consider only 
such orbifolds with small volume (large $\nu_1$) then the
dominant contribution comes from the first term in the right hand 
side in (\ref{eq:Weyl}). 
If we further assume that the eigenmodes have the same 
pseudo-random property as those of CH manifolds\cite{AS93}
then we can readily estimate the angular powers. As shown in figure 
\ref{fig:TLorbifolds} the suppression of the large-angle power 
for small orbifold models is still prominent 
in the case of nearly flat geometry. 
\BF[tp]
\centerline{\psfig{figure=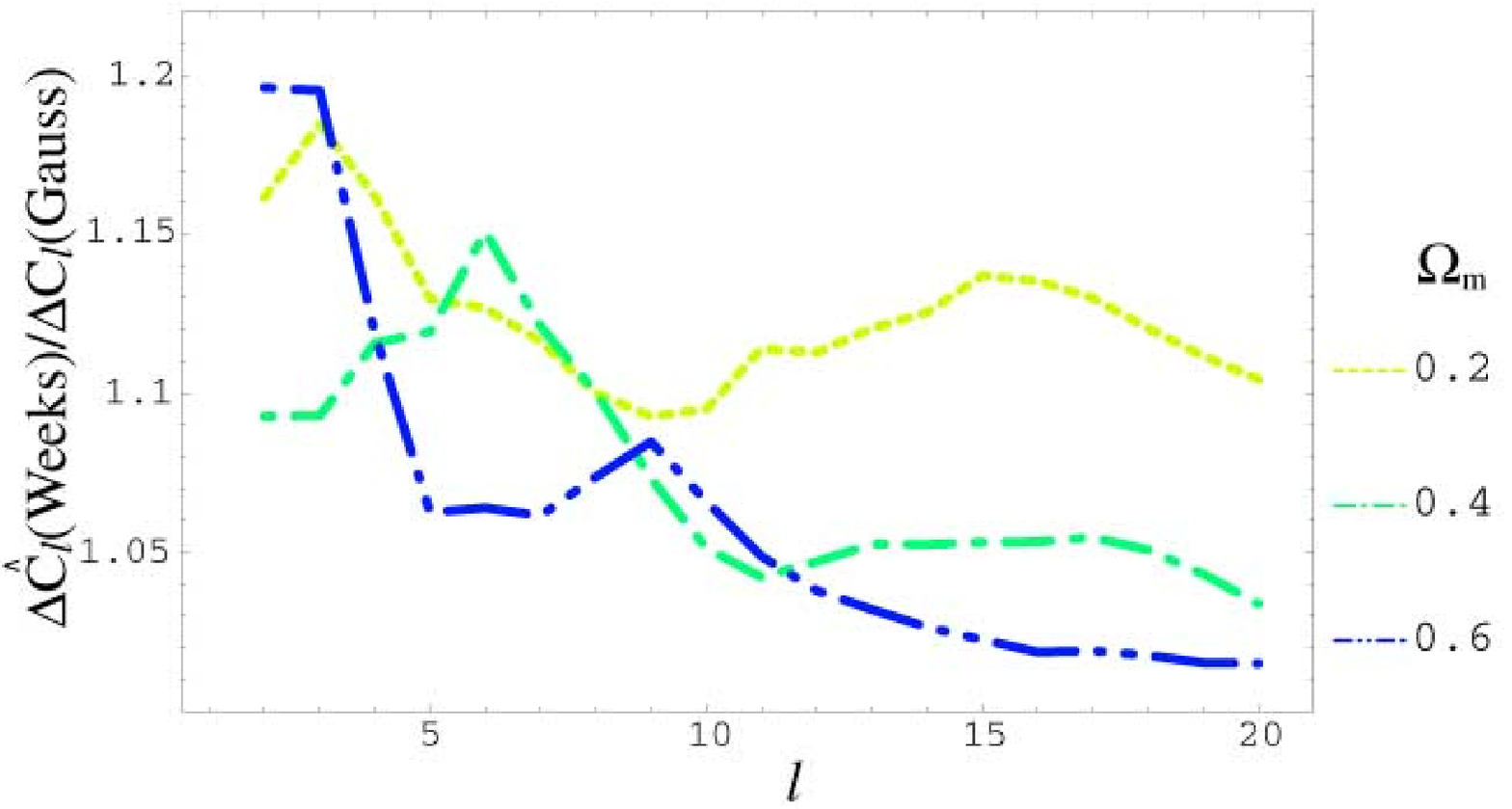,width=14cm}}
\caption{Fractional uncertainty in the angular power 
 $\Delta \hat{C}_l=(\textrm{Var}(\hat{C}_l))^{1/2}/
<\hat{C}_l>$ for the
Weeks models relative to the values for the Gaussian models
$\Delta
C_l(\textrm{Gauss})=(\textrm{Var}(C_l))^{1/2}/<C_l>=(2/2l+1)^{1/2}
$}
\label{fig:cosmicvariance}
\EF
\\
\indent
Another feature of the non-trivial topology is an increase in the
cosmic variance which is attributed to 
the global inhomogeneity and anisotropy of
the background geometry.
If we assume the pseudo-random Gaussianity of the expansion
coefficients $\xi_{\nu l m}$ as observed in the smallest CH
manifolds\cite{Inoue1,Inoue3} then
uncertainty in the power (which may be 
called the \ti{geometric variance}) can be easily estimated. 
For CH models, $a_{lm}$ is 
given by a sum of products of the initial perturbation $\Phi_\nu(0)$
times the expansion coefficients 
\BE
a_{l m}=\sum_{\nu} \Phi_\nu(0) \xi_{\nu l m}F_{\nu l}.
\EE
\label{eq:almCH}
If we fix the values of the 
primordial fluctuations then $a_{l m}$'s  behave as 
if they are random Gaussian numbers and the angular power obeys
the $\chi^2$ distribution. Because the degree of freedom
($k,m$) in the sum is always larger than $2l+1$,
the geometric variance should be always smaller 
than the ``initial variance'' owing to the 
uncertainty in the initial conditions. Therefore, we expect that 
the net cosmic variance 
($\approx $initial variance $+$ geometric variance) is not significantly
greater than the values for the infinite counterpart.
\\
\indent
In order to confirm the validity of the Gaussian assumption for
computing the cosmic variance, firstly, 
we have compared the fractional uncertainty in the angular power 
$(2\le l \le 20)$ of the Weeks models
using only the lowest 33 
numerically computed 
eigenmodes ($k<13$) to those using the 
Gaussian random approximation for the expansion coefficients 
(corresponding to the same eigenvalues). It has turned out that the
errors lie within several per cent 
for models with $\Omega_m=0.2,0.4$ and $0.6$(the relative 
errors are $0.08$ for $l=2$ and $<0.04$ for $l>2$). 
\\
\indent
Next, taking the contribution of higher modes into account,
we have computed the fractional uncertainty in the angular power, 
which can be estimated by using the 
Gaussian approximation for the expansion coefficients (corresponding to
the computed eigenvalues for $k<13$ and the approximated 
eigenvalues obtained from Weyl's asymptotic formula
for $k<60$) for the same parameters and we have compared
the values with those using only the lowest 33 
eigenmodes ($k<13$) and it has found that the errors lie within   
several per cent(however, for high matter density models the systematic
increase in the variance owing to the artificial mode cut-off 
is much significant as the number of modes that contribute to the
sum grows).
As shown in figure \ref{fig:cosmicvariance}, on large angular scales 
the fractional uncertainty in the angular power (for the Weeks models) 
with low matter density increases just 5 to 20 percent relative to  
the one for the Gaussian models. Thus the previous claim that the 
$C_l$'s have large cosmic variances\cite{Bond2} is not correct.
On small angular scales, an increase in the number of 
eigenmodes that contributes to the power leads to a decrease in the 
geometric variance. Consequently, 
the cosmic variance converges to the one for 
the Gaussian model as implied by the central limit theorem. 
\section{Bayesian analysis}
In this section, we study the likelihoods of CH models using the COBE   
data. Although here we only study manifold models, we expect the result
for orbifold models (with the same volume) will not 
grossly change since the statistical
property of eigenmodes are expected to be similar with those of
manifolds\cite{Aurich99}.
\\
\indent
The covariance in the temperature at pixel $i$ and 
pixel $j$ in the sky map is given by
\BE
M_{ij}=<T_i T_j>=\sum_{l}<a_{lm}a_{l'm'}>W_l W_l'Y_{l m}(\hat{n_i})
Y_{l' m'}(\hat{n_j})+<N_i N_j>,
\EE
where $<>$ denotes an ensemble
average taken over all initial conditions, positions
and orientations of the observer,\footnote{Here we assume that we do not
know anything about the position and orientation of the observer. The
covariance is defined for an isotropic and homogeneous ensemble of observers. 
  } $T_i$ represents the
temperature in pixel $i$, $W_l^2$ is the
experimental window function that includes effects of 
beam-smoothing and finite pixel size,
$\hat{n_i}$ is the unit vector towards the center of pixel $i$
and $<N_i N_j>$ is the noise covariance between pixel $i$ and pixel $j$. 
If the fluctuations in the sky form an isotropic Gaussian field 
then the covariance is written as 
\BE
M_{ij}=\f{1}{4 \pi}\sum_{l}(2l +1) W_l^2 C_l P_l
(\hat{n_i}\cdot \hat{n_j})+<N_i N_j>,
\label{eq:M}
\EE
where $P_l$ is the Legendre function. 
Assuming a uniform prior distribution for a set of cosmological 
parameters, the probability distribution function of a power spectrum $C_l$
is given by
\BE
\Lambda(C_l|\vec{T})\propto\f{1}{\det^{1/2}M(C_l) } 
\exp \Biggl (\f{1}{2}\vec{T}^T\cdot M^{-1}(C_l) \cdot \vec{T}\Biggr), 
\label{eq:LL}
\EE
where $\vec{T}$ denotes an array of the data of the temperature at
pixels. 
\\
\indent
In the following analysis, we use the inverse-noise-variance-weighted
average map of the 53A,53B,90A and 90B  
COBE-DMR channels. To remove the emission from the galactic
plane, we use the extended
galactic cut (in galactic coordinates).\cite{Banday}
After the galactic cut, best-fit monopole and dipole are removed  
using the least-square method.
To achieve efficient analysis in computation,  
we further compress the data at ``resolution
6'' $(2.6^o)^2$  pixels  into one at ``resolution
5'' $(5.2^o)^2$ pixels for which there are 1536 pixels in the
celestial sphere and 924 pixels surviving the extended galactic cut.
The window function is given by $W_l=G_l F_l$ where $F_l$ are the Legendre
coefficients for the DMR beam pattern\cite{Lineweaver} 
and $G_l$ are the Legendre
coefficients for a circular top-hat function with area equal to the
pixel area which account for the pixel smoothing effect 
(the effect of the finite pixel size is non-negligible 
since the COBE-DMR beam FWHM is comparable to the size of ``resolution
5'' pixels).\cite{Hinshaw}
To account for the fact that we do not have useful information
about the monopole and dipole anisotropy, 
we set $C_0=C_1=100 mK^2$ which renders the likelihood insensitive to
monopole and dipole moments of several $mK$. We also assume
that the noise in the pixels is uncorrelated which is found to be
a good approximation.\cite{Tegmark-Bunn}
\\
\indent
First of all, we set the initial condition as 
$\Phi^2_\nu(0)\!\propto\!1 /(\nu (\nu^2+1))$ in order to 
approximately estimate the effect of the 
suppression in the large-angle power 
owing to the non-trivial topology (here we do not consider $\Phi_\nu(0)$
as random numbers). Then the fluctuations form a pseudo-Gaussian 
random field assuming pseudo-Gaussianity of the eigenmodes.
\BF[t]
\centerline{\psfig{figure=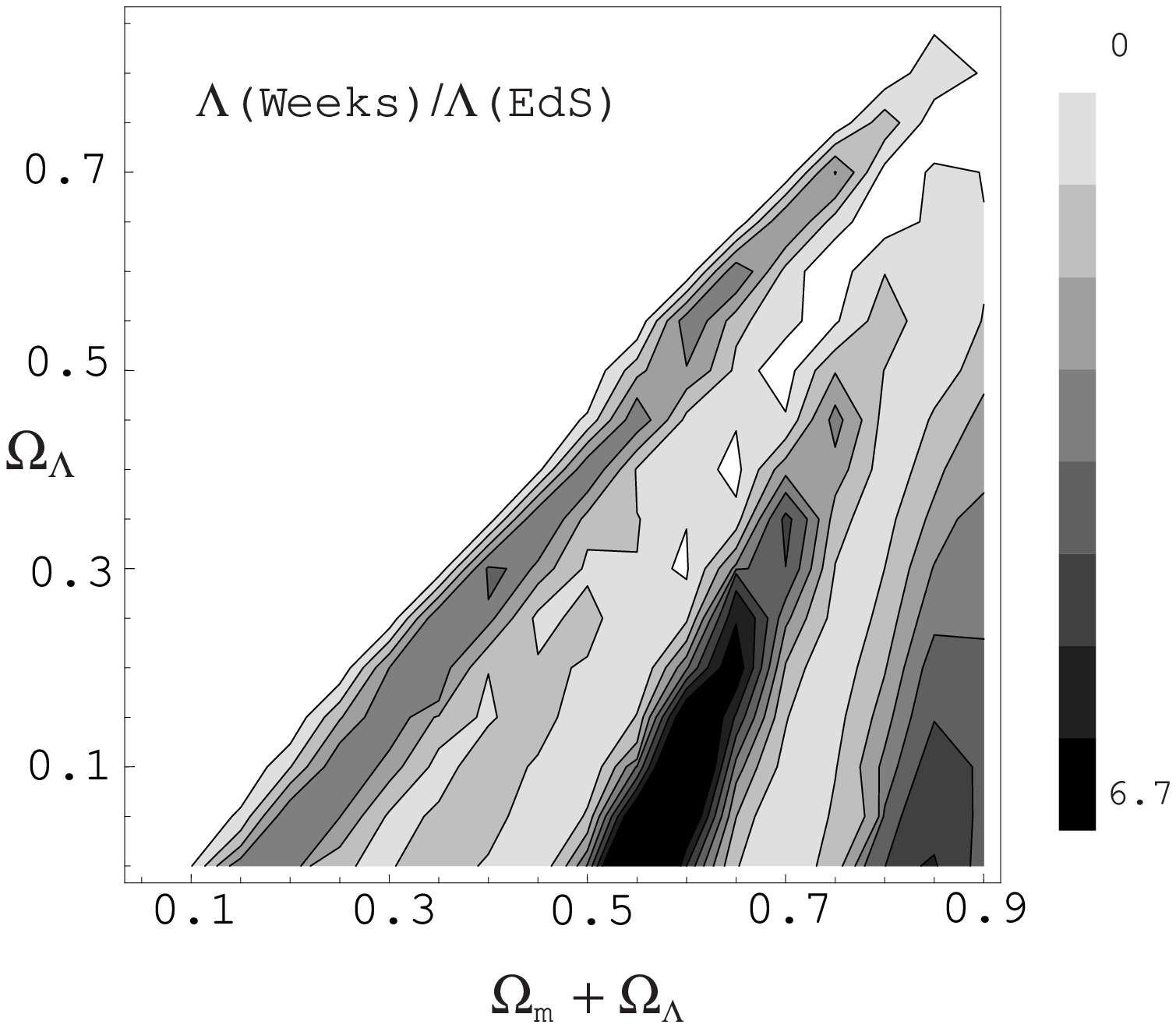,width=8cm}
\psfig{figure=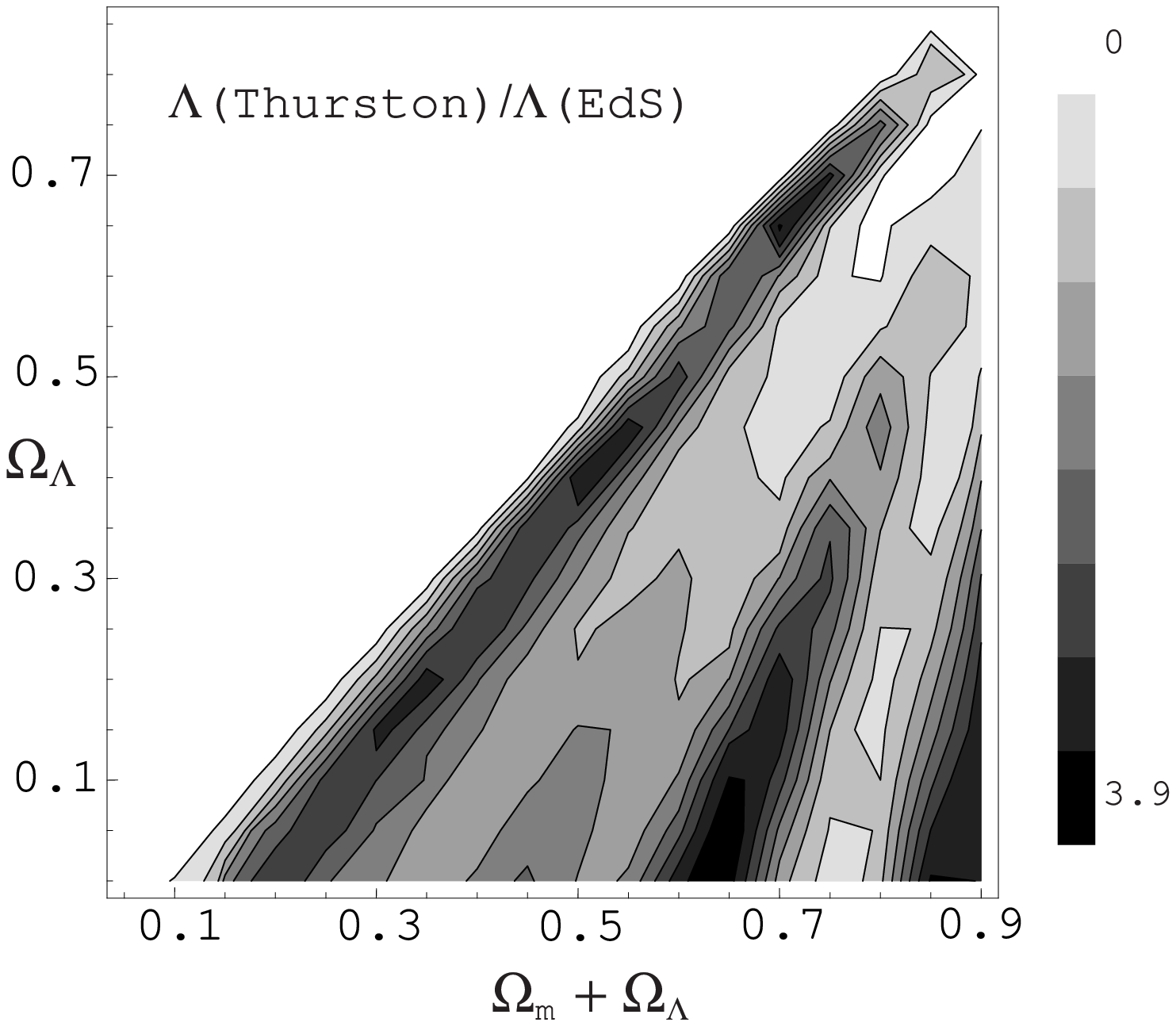,width=7.4cm}}
\caption{Plots of the ratio of likelihoods (assuming that the initial 
condition is fixed) for two smallest 
CH models (Weeks and Thurston) to a likelihood for the Einstein-de
Sitter model
($\Omega_m=1.0$) with scale invariant initial spectrum ($n\!=\!1$).
All the likelihoods are marginalized over the normalization of the 
power. Here we have assumed 
$\Phi^2_\nu(0)\!\propto\!1 /(\nu (\nu^2+1))$ for CH models.} 
\label{fig:LLWT}
\EF
\BF[t]
\centerline{\psfig{figure=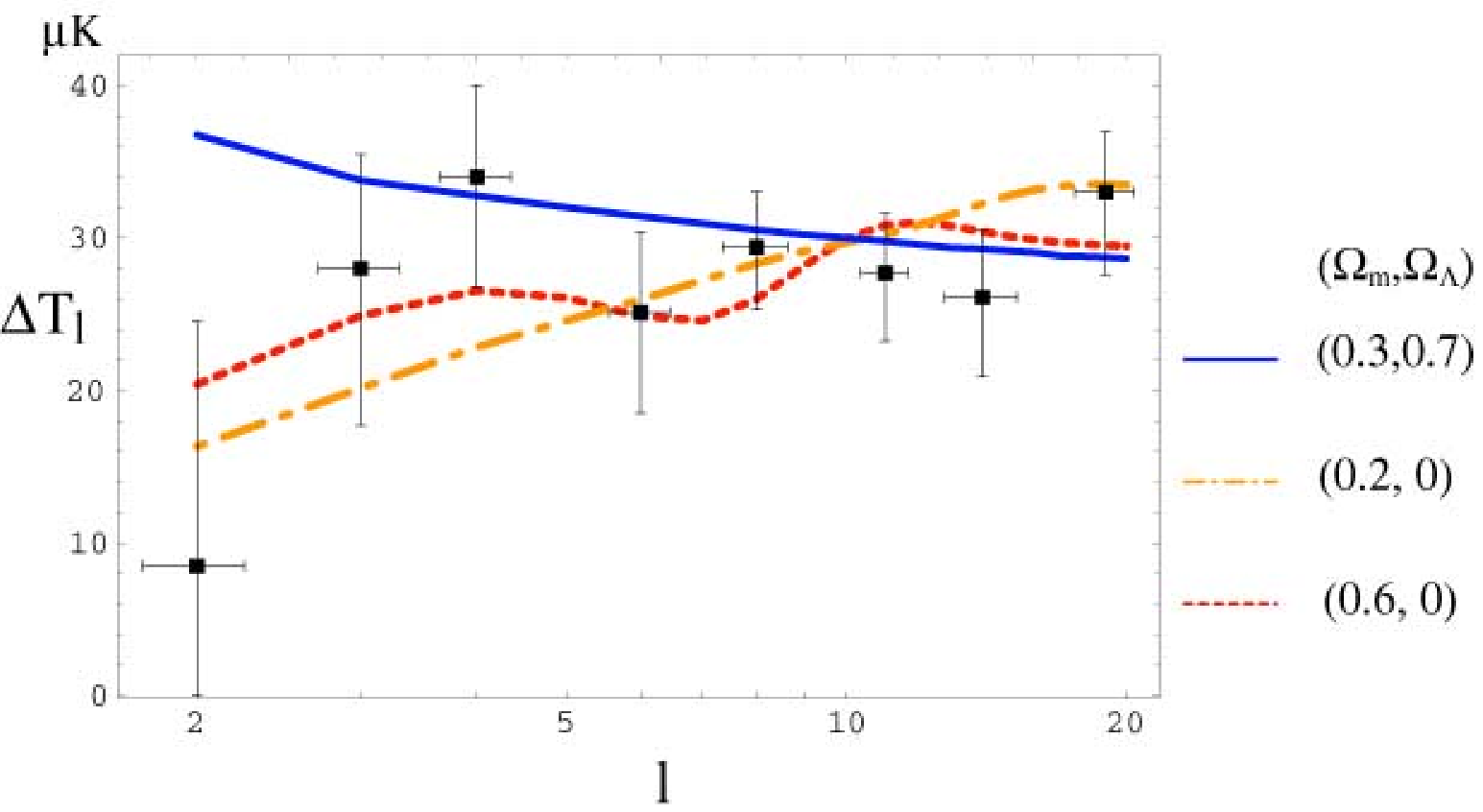,width=15cm}}
\caption{Plots of angular power spectrum 
$\Delta T_l=(l(l+1)\hat{C}_l/2 \pi)^{1/2}$ of the COBE-DMR data (box)
\cite{Tegmark} in comparison with those for the Thurston models 
with $\Omega_m=0.2,0.6$ and
a flat $\Lambda$ model with $\Omega_m=0.3$ and $\Omega_\Lambda=0.7$.}
\label{fig:COBE}
\EF
\BF[t]
\centerline{\psfig{figure=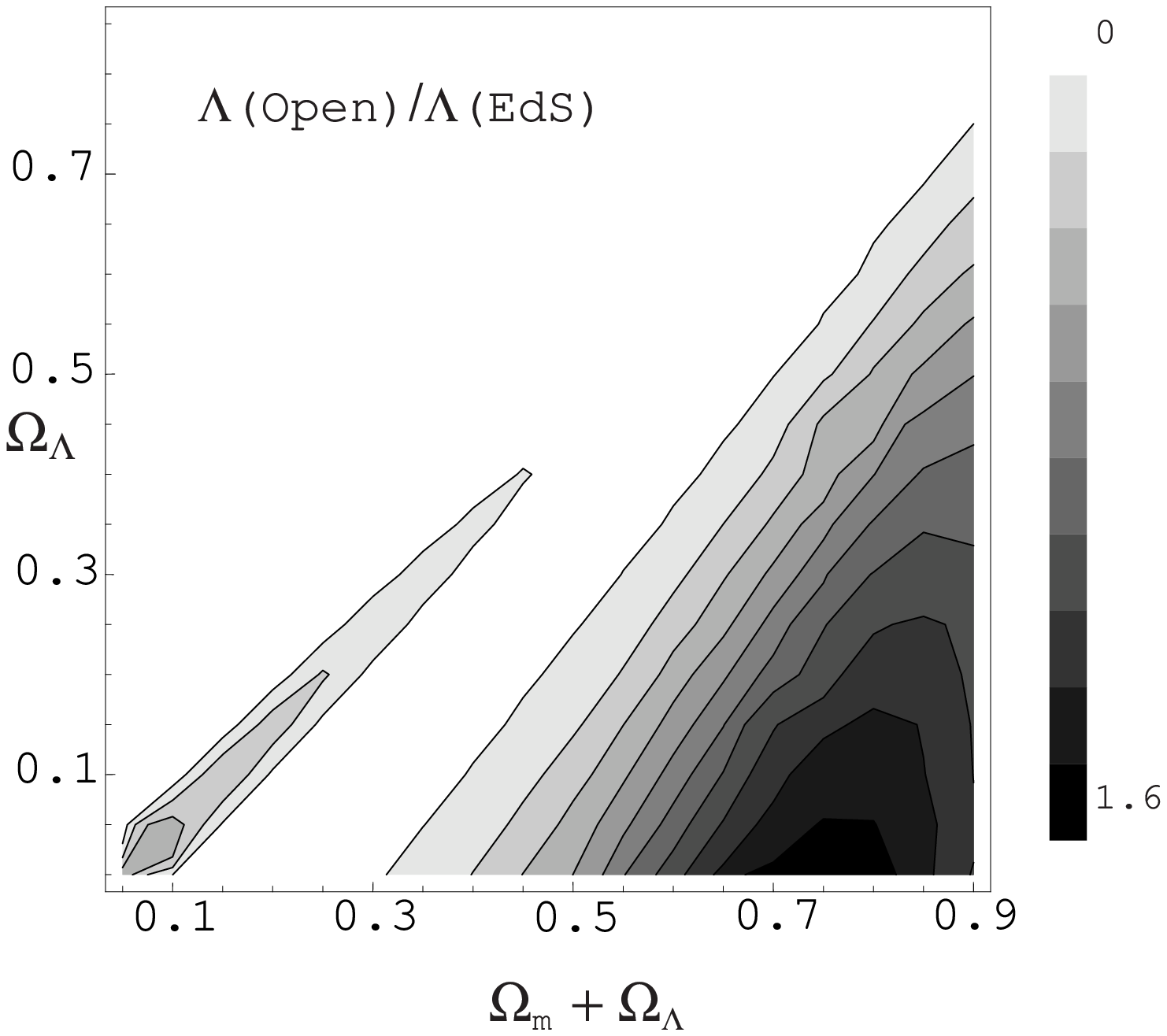,width=8.5cm}
\psfig{figure=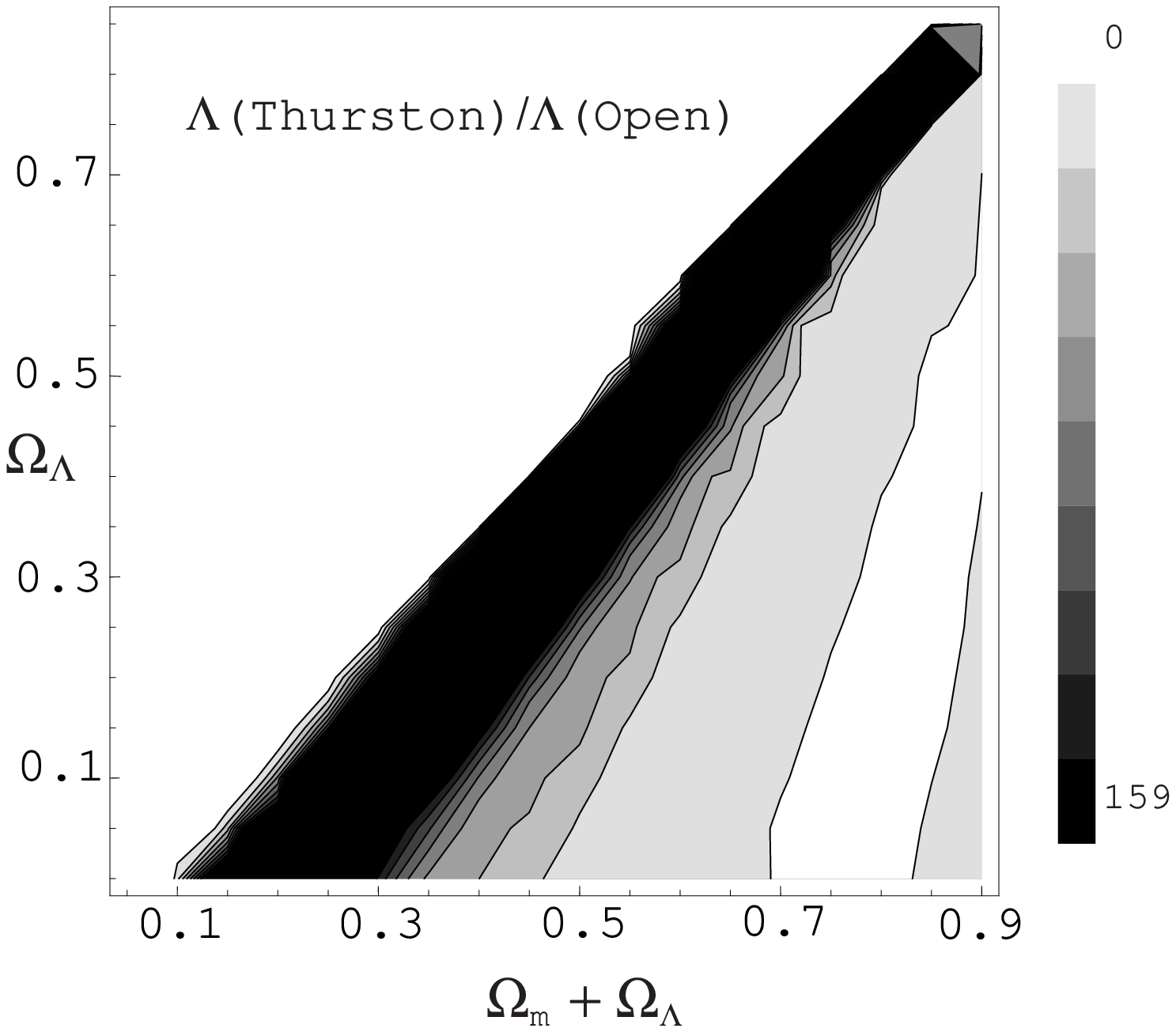,width=8.5cm}}
\caption{Plots of the ratio of likelihoods 
for the infinite hyperbolic ``open'' models 
($n\!=\!1$) to one for the Einstein-de Sitter model ($n\!=\!1$) (left)
and likelihoods for the Thurston models relative to one
for infinite hyperbolic models (right).
All the likelihoods are marginalized over the normalization of the 
power. Here we have assumed 
$\Phi^2_\nu(0)\!\propto\! 1/(\nu (\nu^2+1))$ for the Thurston models.
The slight improvement in the likelihood of infinite hyperbolic models
with $\Omega_m<0.1$ is caused by the absence of the supercurvature
modes. For the Thurston models with $\Omega_m=0.1 \sim 0.3$ the 
likelihoods are significantly improved.
} 
\label{fig:LLTO}
\EF
As shown in figure \ref{fig:LLWT} for a wide 
range of parameters ($\Omega_m+\Omega_\Lambda>0.1$)
the likelihoods for the smallest 
CH manifold models (Weeks and Thurston)  
are better than one for the Einstein-de Sitter model
with scale-invariant spectrum ($n=1$) where $\Delta T_l=
(l(l+1)C_l/2 \pi)^{1/2}$ is almost constant in $l$.
One can see the better fits to the COBE data 
for three parameter regions:1.
$\Omega_m= 0.5 \sim 0.7$ with small $\Omega_\Lambda$ for which
the angular power  
is peaked at $l\sim 4$ which corresponds
to the first ISW ridge of the transfer function;
2. $\Omega_m=0.85\sim0.9$ where the angular scale $l\sim 4$ which
corresponds to the SW ridge;3. $\Omega_m\sim0.2$
where the slope of the power on large angular scales $l<10$
fits well with the data. 
\\
\indent
For low matter density models with infinite volume(that is simply
connected) the ISW effect leads to an excess power on large angular scales.
Therefore the fit to the COBE data is not good because of the 
low quadrapole moment in the data(figure \ref{fig:COBE}).
In contrast, for small CH models, as we have seen, 
the excess power owing to the ISW effect is mitigated by a 
suppression owing to the mode-cutoff of the eigenmodes.  
Therefore, likelihoods for small CH models with low matter density 
are significantly improved compared with the infinite counterparts
(figure \ref{fig:LLTO}). 
As the volume increases the likelihoods converge
to those of the infinite counterparts although the convergence rate 
depends on cosmological parameters.
One can see in figure \ref{fig:LL27CHM} that 
the conspicuous difference for $\Omega_m\!=\!0.1-0.3$ still persists
for volume$\sim 6$ whereas such difference is not observed for
nearly flat cases ($\Omega_{tot}\!=\!\Omega_m+\Omega_\Lambda\!=\!0.9$).
Roughly speaking, the difference in the power
depends on the number $N$ of the copies of the
fundamental domain inside the observable region at present.
\BF[t]
\centerline{\psfig{figure=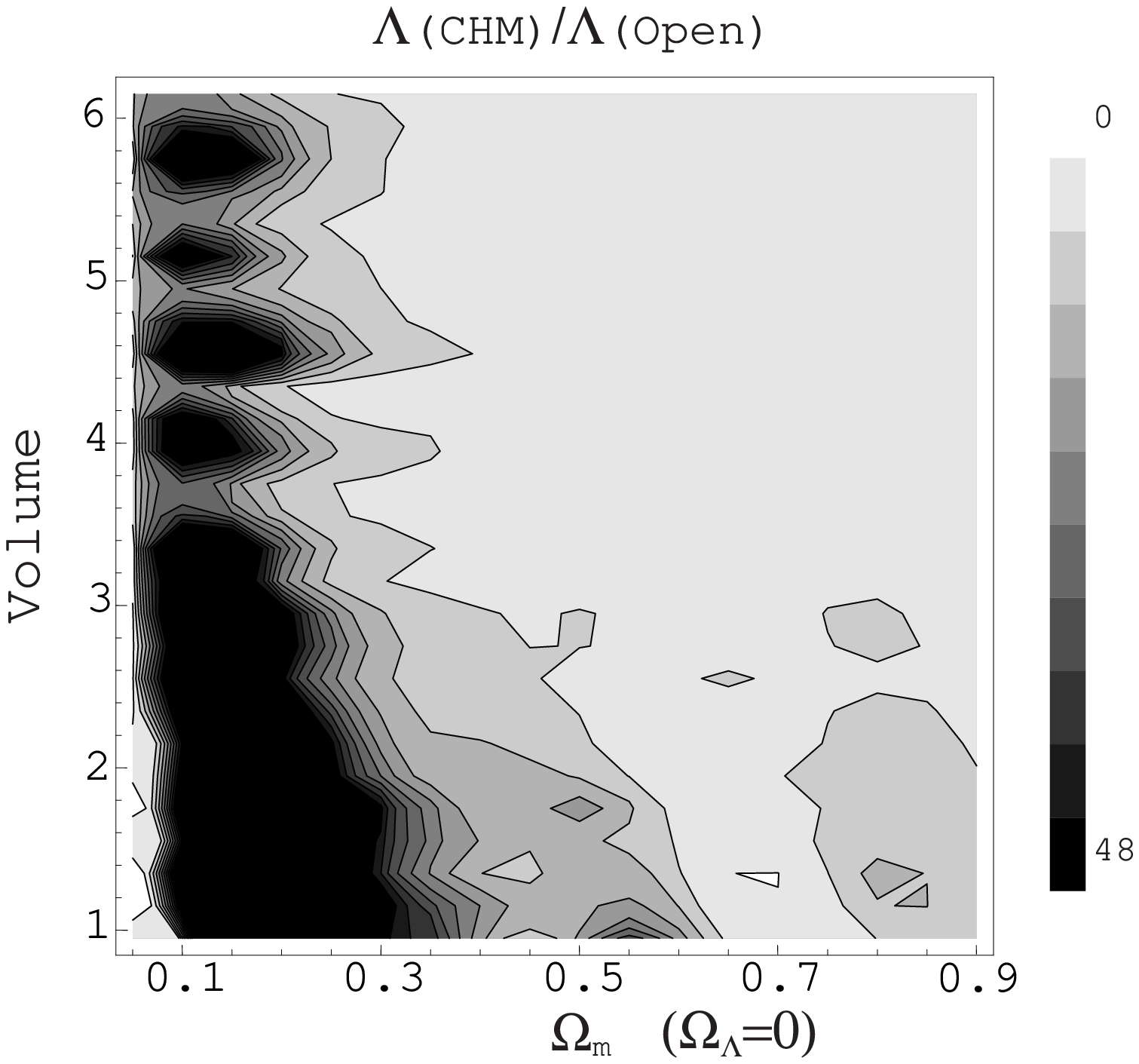,width=8.5cm}
\psfig{figure=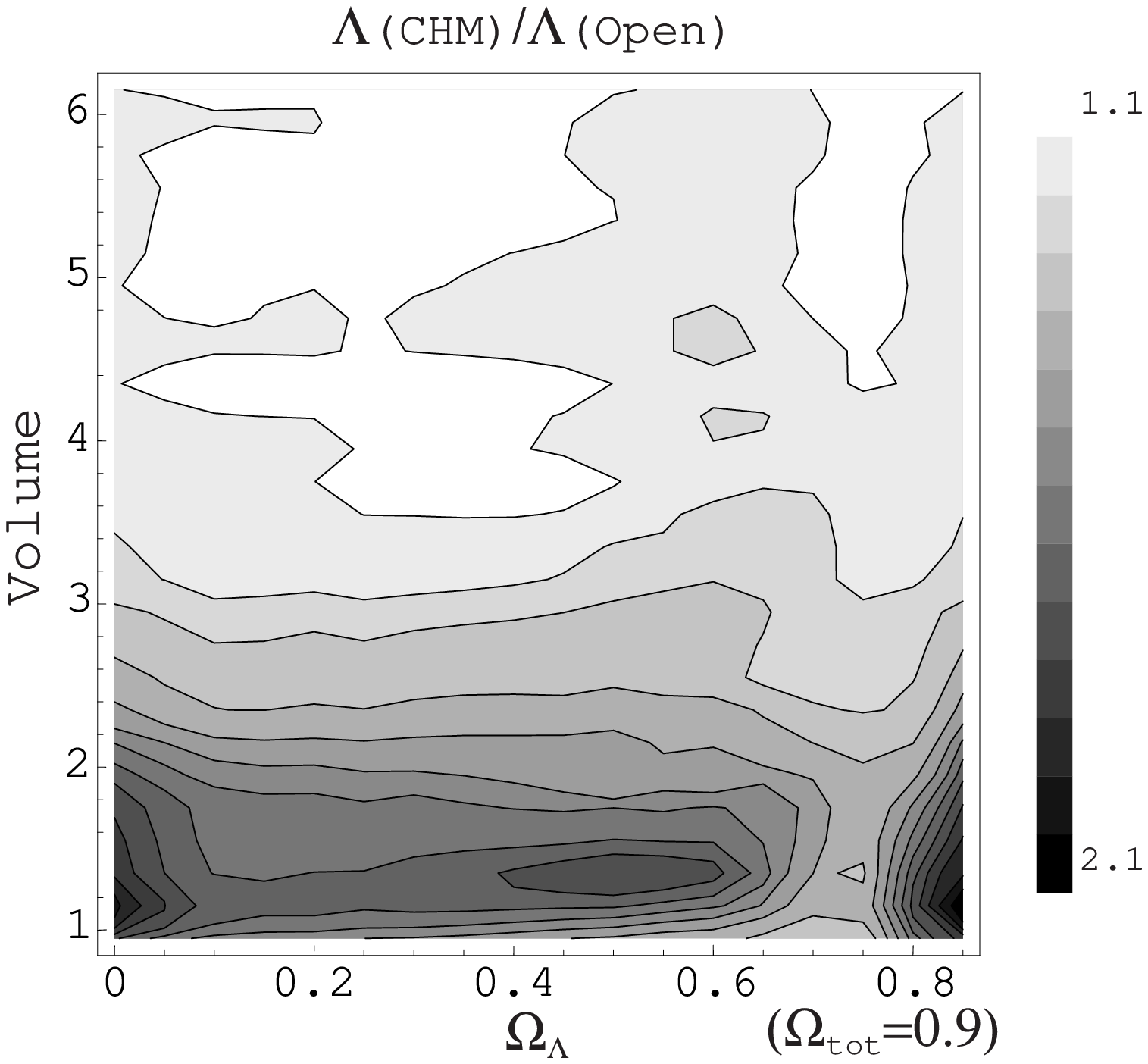,width=8.5cm}}
\caption{Plots of the ratio of approximated likelihoods for 27  
CH models with volume ($0.94-6.15$) 
to a likelihood for the infinite hyperbolic counterparts. 
All the likelihoods are marginalized over the normalization of the 
power. Here we have assumed 
$\Phi^2_\nu(0)\!\propto\! 1/(\nu (\nu^2+1))$ for CH models.} 
\label{fig:LL27CHM}
\EF
\BF
\centerline{\psfig{figure=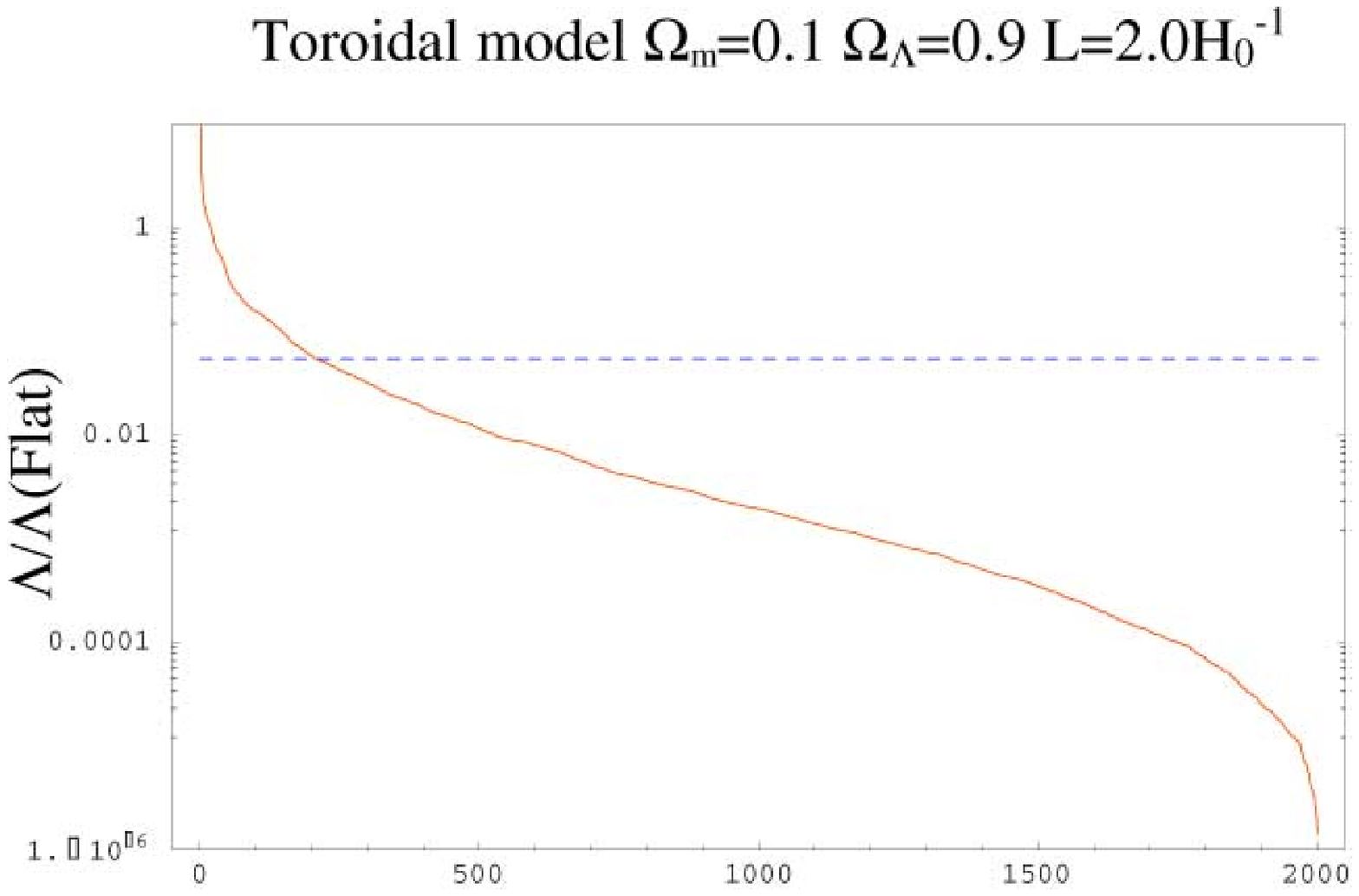,width=11cm}}
\centerline{\psfig{figure=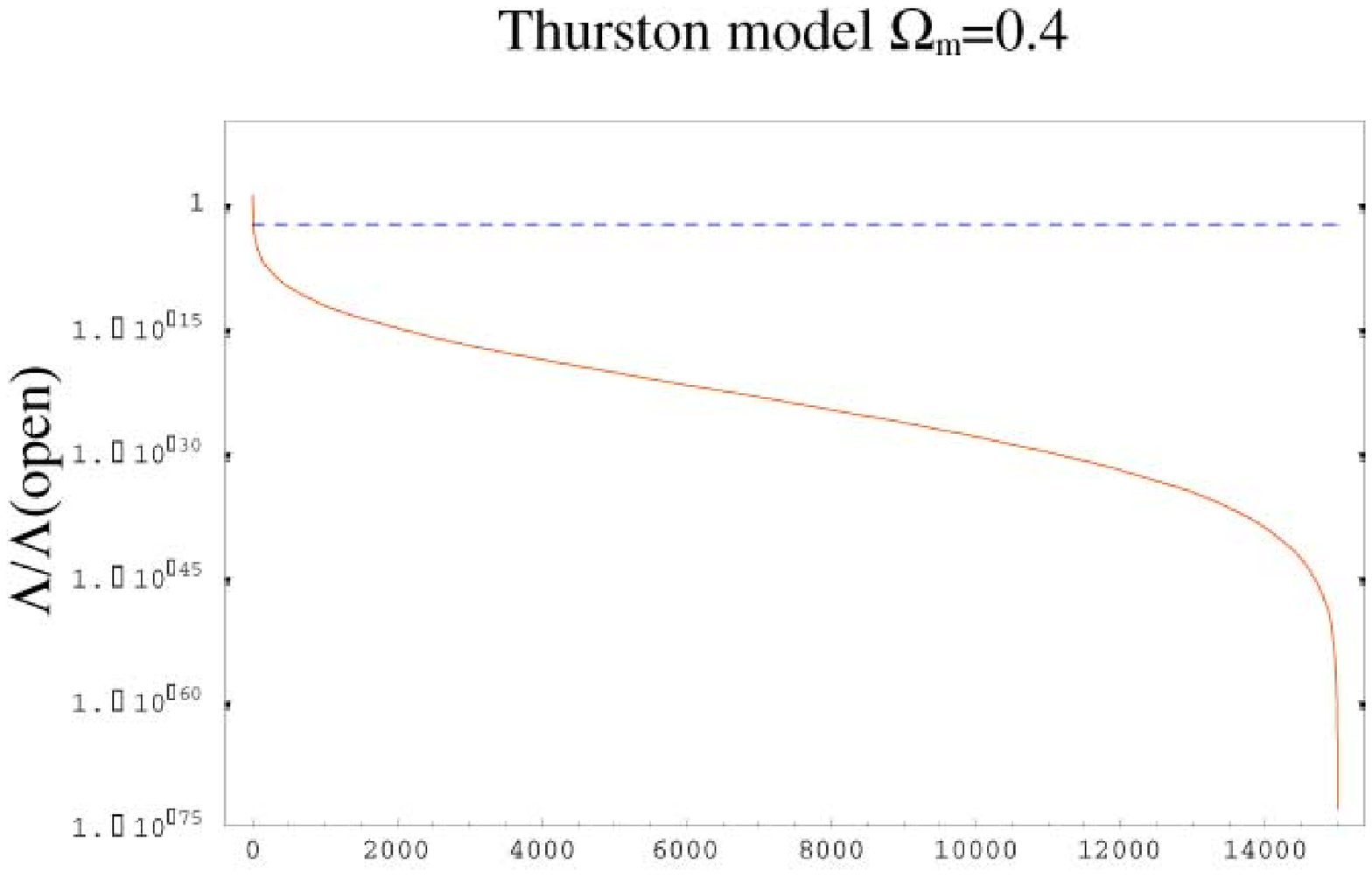,width=11.5cm}}
\caption{Plots of the ratio of ``rigorous'' likelihoods 
incorporating the effect of inhomogeneity and anisotropy of the
background geometry for two compact multiply
connected models relative to one for the 
infinite counterparts in ascending 
order(full curve) for 2000 random realizations of orientation for a 
closed flat toroidal (where the Dirichlet fundamental domain is a cube 
with sides $L=2.0 H_0^{-1}$) model with
$\Omega_m=0.1,\Omega_\Lambda=0.9$ (top) and for  
1500 random realizations of position and 10 realizations of
orientation for a Thurston model
with $\Omega_m=0.4$ (bottom).
The dashed lines denote the ensemble averaged values.
The number $N$ of the copies of fundamental domain inside
the observable region is 64.7 for the compact
flat toroidal model and 72.3 for the Thurston model.}
\label{fig:LLND}
\EF
\BF
\centerline{\psfig{figure=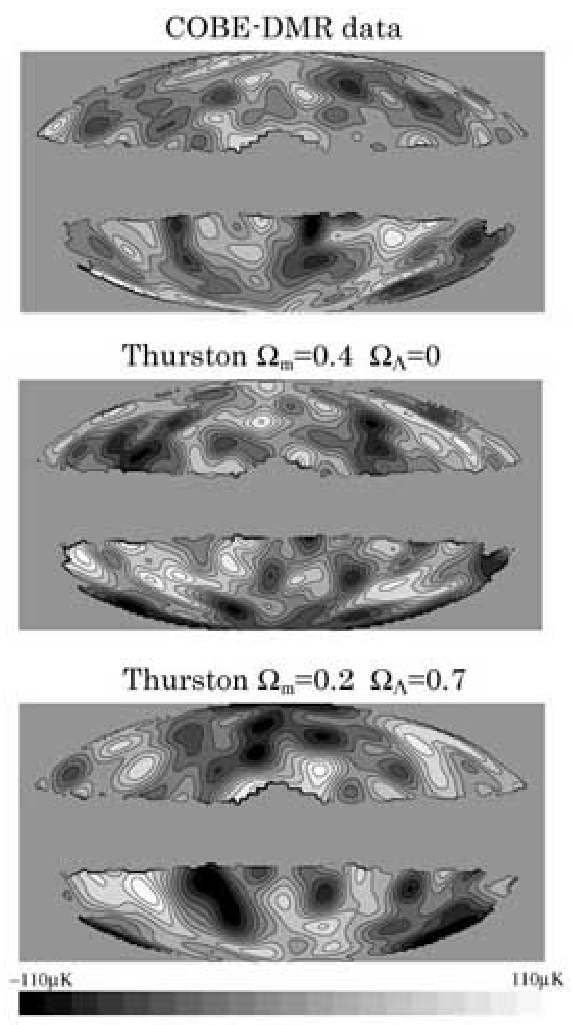,width=10cm}}
\caption{The top map shows the inverse-noise-variance-weighted 
averaged map of the COBE-DMR data at 53GHz and 90GHz after Wiener
filtering assuming an angular power spectrum (averaged over the
position of the observer) for a Thurston model 
with $\Omega_m=0.4$. Each of the mid and the bottom map shows a random 
realization of the CMB anisotropy (convolved with the COBE-DMR beam)
at a position that gives a best-fit to the COBE-DMR data. For all the
maps the ``extended galactic cut'' proposed by the 
COBE-DMR team was used and both the monopole and dipole were removed.}
\label{fig:skymap}
\EF
\\
\indent
Next, we consider the effect of the non-diagonal elements which we
have neglected so far. 
The likelihood for a homogeneous and isotropic ensemble is obtained by
marginalizing the likelihoods all over the positions $\x_{obs}$ 
and the orientations $\alpha$ of the observer,
\BE
\Lambda=\int\!\!\sqrt{g}~ d\x_{obs} d\alpha \Lambda(\x_{obs},\alpha),
\EE
where we assume a constant distribution for the volume elements 
$\sqrt{g} d\x_{obs}$ and $d\alpha$ 
of a Lie group $SO(3)$ with a Haar measure.   
Assuming that the initial fluctuations are Gaussian with 
Harrison-Zel'dovich spectrum ($n=1$), the 
likelihood $\Lambda(\x_{obs},\alpha))$ is given by 
(\ref{eq:M}) and (\ref{eq:LL}) where 
$<a_{lm}a_{l'm'}>$ is written as
\BE
<a_{lm}a_{l'm'}> \propto \f{1}{\nu(\nu^2+1)}
\xi_{\nu l m} \xi_{\nu l' m'}F_{\nu l} F_{\nu l'},
\EE
where $<>$ denotes an ensemble average
taken over the initial condition 
and $F_{\nu l}$ describes contribution from 
the OSW effect and the ISW effect, respectively.
Note that $\xi_{\nu l m}$'s are functions of $\x_{obs}$ and $\alpha$.
In order to compute likelihoods, we use a compressed 
data at ``resolution 3''$(20.4^o)^2$
pixels in galactic coordinates for which there are 60 pixels surviving the 
extended galactic cut for efficient analysis in computation. Although the 
information of fluctuations on small angular scales $l>10$ is lost, we
expect that they still provide us sufficient information for
discriminating the effect of the non-trivial topology which is
manifest on large-angular scales. 
\\
\indent
For the Thurston model, it has turned out that only 0.09
percent of the total of 1500 positions with 10 orientations are larger
than the mean value. We have also computed likelihoods 
for 5100 positions with 40 orientations. Then the 
percentage has reduced to 0.02 from 0.09. The ratio
of the likelihood marginalized over 5100 positions and 40 orientations
to the likelihood of the infinite hyperbolic model with $\Omega_m\!=\!0.4$
is $\textrm{Log}_{10}(\Lambda/\Lambda(\textrm{open}))=-1.4$ and 
the maximum value is 
$\textrm{Log}_{10}(\Lambda(\max)/\Lambda(\textrm{open}))=3.6$.
For 17 cases out of 204000 realizations, the likelihoods are much
better than one for the infinite counterpart.
Although 99.98 percent choices of position and
orientation are ruled out, the likelihood for the remaining choices  
is approximately 4000 times larger than that for the infinite counterpart
which boosts the probability of having skymaps consistent with the data.
It has turned out that the positions that give a better fit
to the data are scattered in the manifold and do not coincide with the
point where the injectivity radius is locally maximal(=the center of the
Dirichlet domain). This suggests that we are accidentally 
put at a certain place with a certain orientation. 
In other words, the observed sky 
gives us partial information about
the position and orientation of the observer in the manifold
although it is not enough to determine the values uniquely.
The best-fit quadrupole normalization is $Q=(5C_2/4\pi)^{1/2}
=0.022\pm 0.0024 mK$
for 50 choices of position and orientation of the observer 
that satisfy $\Lambda/\Lambda(\textrm{open})>0.03$
while $Q=0.027\pm 0.0052 mK$ 
for a total of 15000 realizations.
For ``bad'' choices the best-fit normalization is somewhat high
since it gives a large cosmic variance whereas the normalization 
is much lowered for ``good'' choices. One can see in figure \ref{fig:skymap}
that a random realization of the skymap for the Thurston models does not 
appear grossly inconsistent with the COBE-DMR data. It turns out that 
\ti{the statistically averaged anisotropic correlation pattern
depends sensitively on the position of the observer}.
\\
\indent
The likelihood analyses in \cite{Bond1,Bond2} are based on correlations 
for a particular choice of position $P$ where the 
injectivity radius is locally maximal with 24 orientations.  
It should be emphasized that the 
choice of $P$ as an observing point is very special one. For instance 
it is the center of tetrahedral and Z2 symmetry of the Dirichlet domain 
of the Thurston manifold 
although they are not belonging to 
the symmetry of the manifold.\cite{Inoue3} 
In mathematical literature it is standard to choose $P$ 
as the base point which belongs to a ``thick'' part of the 
manifold since one can expect many symmetries. 
However, considering such a 
special point as the place of the observer cannot 
be verified since it is inconsistent with the 
Copernican principle. Because any CH models are globally inhomogeneous,
one should compare fluctuation patterns expected 
at \ti{every} place with \ti{every} orientation. 
\\
\indent
To see the influence of global inhomogeneity, it is 
illustrative to compare the relative likelihoods(to the infinite
counterpart) of the Thurston 
model with those of compact flat toroidal model (obtained by 
gluing opposite faces of a cube by three translations) which is
globally homogeneous (i.e. every action of the discrete isometry
group is a Clifford transformation). As shown in figure 10, for 
the toroidal model(that has approximately the same 
proportion to the currently observable region in size in comparison with
the Thurston model), 
dependency on choices of 
position of the observer is less significant. 
Therefore one does not need a number of 
realizations for estimating the likelihood. 
In contrast, for CH models, one needs a sufficient
number of realizations since the proportion of 
choices of position and orientation of the observer that give a better
fit to the data is considerably small. 
Thus if one treats CH models
like the toroidal model \cite{Bond1,Bond2}, one gets 
misleadingly small values of the likelihoods.
\\
\indent
Although our result is based on the numerical computation,
it is a natural result if one knows the pseudo-random
behavior of eigenmodes of CH spaces. 
For each choice of position and orientation of the
observer, a set of expansion coefficients $\xi_{\nu l m}$ of eigenmodes 
is uniquely determined (except for the phase factor), 
which corresponds to a ``realization'' of independent random 
Gaussian numbers. By taking an average over the position and the
orientation, the non-diagonal terms proportional to 
$<\xi_{\nu lm}\xi_{\nu l'm'}>, l \ne l', m \ne m'$ vanish. In other
words, a set of anisotropic patterns all over the place in a CH
space comprises an almost isotropic random field. Consider two realizations 
$A$ and $B$ of such an isotropic random field. The chance you would
get an almost similar fluctuation pattern for $A$ and $B$ would be 
very low but we do have such an occasion. Similarly, the likelihood at a
particular position with a certain orientation is usually very low
but there \ti{are} cases for which the likelihoods are considerably high.
Thus we conclude that the COBE constraints on small CH models are
less stringent as long as the 
Gaussian pseudo-randomness of the eigenmodes holds. 
\section{Summary}
In this paper, we have explored the CMB anisotropy in 
small CH models with or 
without the cosmological constant.
Assuming adiabatic initial perturbation with scale-invariant 
spectrum ($n=1$), the angular power owing to the OSW 
effect suffers a prominent 
suppression since fluctuations beyond the size of the fundamental domain 
at the last scattering are strongly suppressed. 
However, for low matter density models, the suppression in the
large-angle power is less stringent because of the significant 
contribution from the ISW effect caused by the decay of the 
gravitational potential at the $\Lambda$ or curvature dominant epoch.
A slight suppression in the 
large-angle power in such models explains rather
naturally the observed anomalously low quadrupole which is
incompatible with the prediction of the standard
Friedmann-Robertson-Walker models. 
\\
\indent
As we have seen, the likelihood of CH models (assuming
pseudo-Gaussianity of eigenmodes)
depends sensitively on the choice of orientation and position 
of the observer. Because the likelihood marginalized over the  
orientation and position (assuming equal probability
for each choice) is comparable to the value of the infinite
counterpart, we conclude that constraints on CH models are less stringent.
It should be emphasized that the dependence of the likelihood 
on the position of the observer is of crucial importance which has been 
ignored by previous literature. Closed multiply connected constantly
curved 3-spaces that are globally 
homogeneous are limited to some spherical spaces and
flat 3-tori.\cite{Wolf67} For ``bad'' choices of the position and
orientation, the best-fit amplitude tends to have a large value 
(allowing a large variance), leading to misleadingly stringent
constraints on the models. Surprisingly, the statistically averaged 
anisotropy of the correlation seems to disappear 
if marginalized all over the place which is related to the 
pseudo-random property of eigenmodes.
\\
\indent
Even in the case of nearly flat geometry, 
the signature of the non-trivial topology is still 
prominent if the space is sufficiently small compared with the 
observable region at present. If we allow orbifold models
then the volume can be much smaller than manifold models. 
We have seen that a slight suppression in the large-angle
power is still prominent for orbifold models with volume $0.01-0.1 R^3$ 
for $\Omega_m=0.2$ and $\Omega_\Lambda=0.7$ which is consistent 
with the result in.\cite{Aurich00}
However, the existence of singularities (where the 
curvature diverges) may cause some problems for any orbifold models.
If plane-like singularities were present(e.g. tetrahedral orbifolds)
astronomical objects with peculiar velocity would
easily collide with the plane  
(we may call such a model as a ``billiard universe''). 
On the other hand, the observational effects 
caused by the presence of fixed lines or 
``strings'' might be less prominent and much safer.  
\\
\\
\section*{Acknowledgments}
I would like to thank J.R. Weeks and A. Reid for 
sharing with me their expertise in topology and geometry of 
3-manifolds and 3-orbifolds.  I would also like to thank N. Sugiyama and 
A.J. Banday for their helpful suggestions and comments on the data
analysis using the CMB-DMR data.
The numerical computation in this work was carried out  
at the Data Processing Center in Kyoto University and 
Yukawa Institute Computer Facility. 
K.T. Inoue is supported by JSPS Research Fellowships 
for Young Scientists, and this work is supported partially by 
Grant-in-Aid for Scientific Research Fund (No.9809834).


\end{document}